# Chemistry of Dark Clouds: Databases, Networks, and Models

Marcelino Agúndez[†] and Valentine Wakelam[*,‡]

[†]Univ. Bordeaux, LAB, UMR 5804, F-33270 Floirac, France

[‡]CNRS, LAB, UMR 5804, F-33270 Floirac, France

CONTENTS

## 1. INTRODUCTION

The introduction of chemistry into astronomy is relatively recent and can be dated back to about half a century ago, when molecules were discovered in space. In the 1930s it was already known that some molecules such as CH, CN, $C_2$, and TiO are present in the atmosphere of the sun and other stars, as identified through broad absorption features in the visible spectrum.[1] Some stellar spectra also showed narrow absorption features of interstellar origin, some of which were known to arise from ionized atoms such as $Ca^{2+}$ or $Na^+$,[2,3] while some others were later assigned to molecules such as CH or CN.[4,5] Still it was not clear how such molecules could survive the harsh interstellar conditions, but by the early 1940s there was enough evidence on the presence of at least three molecules, CH, CN, and $CH^+$, in the interstellar medium.[6,7] It was not, however, until the 1960s and early 1970s that the development of radioastronomy brought a boom in detections of interstellar molecules (the very first ones being OH, $NH_3$, $H_2O$, $H_2CO$, and CO[8–12]) and paved the road for the birth of astrochemistry.

Cold dense clouds, also known as dark clouds or sometimes simply dense clouds, were first observed by William Herschel as holes in the heavens left by stars[13] and later identified by Edward E. Barnard as obscuring bodies between background stars and us.[14] They are mostly composed of gas-phase molecular material, although their opaque optical appearance is due to a population of tiny dust particles present at the 1% mass level. They are cold (~10 K), dense ($10^3$–$10^6$ particles/ $cm^3$; the term "dense" is used to distinguish them from diffuse clouds where densities are $<10^3\ cm^{-3}$), and dark (they are opaque to visible and UV radiation). Dark clouds present different shapes, although they usually appear with a filamentary structure at large scales, whereas at small (subparsec[15]) scales they show a nonhomogeneous structure with denser cores (where the density may be as high as $10^6\ cm^{-3}$) which are the sites of the birth of stars (see the review on the subject by Bergin and Tafalla[16]).

The exact definition of cold dense clouds concerning their physical parameters and dynamical status is a matter of debate,[16] probably because they are one of the intermediate stages in the evolutionary process that converts diffuse interstellar matter into stars. Due to turbulence and gravitation, large-scale diffuse clouds will evolve toward giant molecular clouds, and within those clouds, local density enhancements can be identified as dense clouds.[17] Those dense clouds will eventually undergo gravitational collapse to form protostars (housing hot cores) and then stars surrounded by planetary systems. Some dense clouds are sometimes referred to as prestellar cores, which are characterized by an increasing gradient of density up to a few $10^6\ cm^{-3}$ toward the center and central temperatures below 10 K. Those prestellar cores can present some complex radial motions such as oscillations.[18,19] Our view of the physical structure of prestellar cores has been largely improved by the recent observations with the Herschel space telescope.[20–22] Other previously identified dense clouds have revealed the presence of infrared (IR) sources inside these clouds, i.e., embedded young stellar objects.[23] For the sake of simplicity and because we focus on the chemistry in this paper, we will use the definition of cold dense clouds by Mayer and Benson,[24,25] who determined mean physical properties with temperatures of 10 K and densities of a few $10^4\ cm^{-3}$.

Table 1. Molecules Detected in Interstellar and Circumstellar Media (July 2013)[a]

| 2 atoms | | 3 atoms | | 4 atoms | 5 atoms | 6 atoms | 7 atoms | 8 atoms |
|---|---|---|---|---|---|---|---|---|
| $H_2$ | CP | $H_3^+$ | MgCN | $CH_3$ | $NH_3D^+$ | $C_2H_4$ | $CH_3NH_2$ | $CH_3CHNH$ |
| LiH ? | AlO | $CH_2$ | NaCN | $NH_3$ | $CH_4$ | $CH_3OH$ | $CH_3C_2H$ | $CH_2CHCHO$ |
| CH | CS | $NH_2$ | $C_2S$ | $CH_2NH$ | $CH_2NH$ | $CH_3CN$ | $CH_3CHO$ | $NH_2CH_2CN$ |
| $CH^+$ | SiO | $H_2O$ | OCS | $C_2H_2$ | $SiH_4$ | $CH_3NC$ ? | $c$-$C_2H_4O$ | $CH_3COOH$ |
| NH | PN | $H_2O^+$ | $SO_2$ | $H_2CN$ | $CH_3O$ | $CH_2NH$ | $CH_2CHOH$ | $CH_2OHCHO$ |
| OH | NS | $C_2H$ | $c$-$SiC_2$ | $H_2CO$ | $H_2COH^+$ | NHCHCN | $CH_2CHCN$ | $HCOOCH_3$ |
| $OH^+$ | AlF | HCN | SiCN | $H_2CO$ | $c$-$C_3H_2$ | $NH_2CHO$ | $C_6H$ | $CH_3C_3N$ |
| HF | PO | HNC | SiNC | $H_2O_2$ | $H_2C_3$ | $CH_3SH$ | $C_6H^-$ | $CH_2CCHCN$ |
| $C_2$ | SO | HCO | $C_2P$ | $PH_3$ ? | $CH_2CN$ | $C_4H_2$ | $HC_5N$ | $C_6H_2$ |
| CN | $SO^+$ | $HCO^+$ | AlNC | $C_3H$ | HNCNH | $H_2C_4$ | | $H_2C_6$ |
| $CN^-$ | NaCl | $HOC^+$ | KCN | $c$-$C_3H$ | $H_2C_2O$ | $HC_4N$ | 9 atoms | $C_7H$ |
| CO | SiS | $N_2H^+$ | $TiO_2$ | $C_3H^+$ | $NH_2CN$ | $HC_3NH^+$ | $CH_2CHCH_3$ | 10 atoms |
| $CO^+$ | AlCl | HNO | FeCN | $HC_2N$ | HCOOH | $HC_2CHO$ | $CH_3OCH_3$ | $CH_3COCH_3$ |
| $N_2$ | TiO | $HO_2$ | | HNCO | $C_4H$ | $c$-$H_2C_3O$ | $CH_3CH_2OH$ | $OHCH_2CH_2OH$ |
| NO | FeO ? | $H_2S$ | | HCNO | $C_4H^-$ | $C_5H$ | $CH_3CH_2CN$ | $CH_3CH_2CHO$ |
| $CF^+$ | KCl | $H_2Cl^+$ | | HOCN | $HC_3N$ | $C_5N$ | $CH_3CONH_2$ | $CH_3C_5N$ |
| SiH ? | | HCP | | $HCO_2^+$ | $HC_2NC$ | $C_5N^-$ | $CH_3C_4H$ | 11 atoms |
| $O_2$ | | $N_2O$ | | $H_2CS$ | $HNC_3$ | | $C_8H$ | $C_2H_5OCHO$ |
| SH | | AlOH | | $C_3N$ | CNCHO | | $C_8H^-$ | $CH_3COOCH_3$ |
| $SH^+$ | | $CO_2$ | | $C_3N^-$ | $C_5$ | | $HC_7N$ | $CH_3C_6H$ |
| HCl | | $HCS^+$ | | $C_3O$ | | >12 atoms | | $HC_9N$ |
| $HCl^+$ | | $C_2O$ | | HNCS | | $HC_{11}N$ | | 12 atoms |
| SiC | | $C_3$ | | HSCN | | $C_{14}H_{10}^+$ ? | | $C_2H_5OCH_3$ ? |
| SiN | | MgNC | | $c$-$SiC_3$ | | $C_{60}$ | | $C_3H_7CN$ |
| | | | | $C_3S$ | | $C_{60}^+$ | | $C_6H_6$ |
| | | | | | | $C_{70}$ | | |

[a]The table includes different isomers (e.g., HCN and HNC) but not different isotopologues (e.g., $H_2O$ but not HDO). Ions are highlighted in red. c stands for cyclic species. A question mark stands for tentative detection. The table was constructed from The Astrochymist[47] and the Cologne Database for Molecular Spectroscopy[48] Web sites.

Among the first ideas proposed by the early 1970s to explain the formation of the different molecules that were by that epoch being discovered in space there were formation through gas-phase binary reactive collisions, formation on the surface of dust grains, and formation in stellar atmospheres and further ejection to the interstellar medium.[26] This latter explanation was definitively rejected as an important source of interstellar molecules on the basis of the difficult survival of molecules to interstellar ultraviolet photons during such a long journey. On the other hand, the two former mechanisms are still today the most widely accepted and constitute the basis of interstellar chemistry. The pioneering contributions to the subject by Bates and Spitzer,[27] Watson,[28–30] and Herbst and Klemperer[31] identified various key aspects of the chemistry of cold dense clouds:

(1) Hydrogen is mainly in molecular form, for which the only efficient formation mechanism is the recombination of hydrogen atoms on the surface of dust grains.

(2) The chemistry is initiated by cosmic rays which, unlike ultraviolet photons, can penetrate inside the cloud and ionize $H_2$ and He. Once $H_2^+$ is formed, it is immediately converted into $H_3^+$ by the reaction

$$H_2^+ + H_2 \rightarrow H_3^+ + H \qquad (1)$$

which is very exothermic (~1.7 eV) and fast.

(3) The $H_3^+$ ion is the key species that initiates the chain of chemical reactions synthesizing most of the molecules in dark clouds. Molecular hydrogen having a proton affinity of 4.4 eV, i.e., lower than those most of the molecules and atoms, $H_3^+$ behaves as a kind of universal proton donor in reactions of the type

$$H_3^+ + X \rightarrow XH^+ + H_2 \qquad (2)$$

where the protonation of the neutral species X renders it chemically activated ($XH^+$ is generally more reactive than X) so that it can participate more easily in different types of chemical reactions.

(4) The very low kinetic temperatures in interstellar clouds, just a few degrees above the 2.7 K cosmic background in cold dense clouds, imply that only chemical reactions that are exothermic and occur without any activation barrier can play a role. Ion–neutral reactions were identified as the main type of reactions able to meet these conditions and to carry out most of the chemical synthesis.

The theoretical framework proposed by Watson, Herbst, and Klemperer to explain the formation of molecules in cold dense clouds was found to be very satisfactory and is still nowadays valid in its main points. In addition to the gas-phase formation pathways, a special interest in chemical reactions occurring on the surface of dust grains existed from the very beginning of interstellar chemistry, mainly due to the need to explain the formation of the most abundant interstellar molecule, $H_2$ [32,33] (see the review by Vidali[34] in this issue), but also aimed at explaining the presence of saturated molecules such as $H_2O$, $NH_3$, and more complex species in warm molecular clouds within so-called hot cores[35,36] (see the reviews by Garrod and Widicus Weaver,[37] Watanabe,[38] and Cuppen[39] in this issue). A turnaround in the scheme of chemical synthesis in cold interstellar clouds occurred during the 1990s, when laboratory experiments demonstrated that some neutral–neutral reactions, in which at least one of the reactants is a radical, are very rapid at low temperatures. This discovery had important implications for interstellar chemistry,[40] which until then based the buildup of molecules in cold dark clouds on reactions involving ions and did not consider neutral–neutral reactions on the assumption that most of these reactions would have activation barriers making them too slow at low temperatures.

One may get a good idea of the chemistry that takes place in interstellar clouds by taking a look at the type of molecules detected. Table 1 lists the more than 170 molecules detected to date in interstellar and circumstellar clouds. By circumstellar clouds we refer to envelopes around evolved stars, the chemistry of which has certain similarities to that of interstellar clouds. The nature of the molecules detected reflects to a large extent the abundances of chemical elements observed in the diffuse medium (from which the dense clouds are formed), the most abundant after hydrogen being helium (0.085), which is inert and rarely forms molecules, oxygen[41] ($2.56 \times 10^{-4}$), carbon[42] ($1.2 \times 10^{-4}$), and nitrogen[42] ($7.6 \times 10^{-5}$), followed by sulfur[43] ($1.3 \times 10^{-5}$), Mg[44] ($2.4 \times 10^{-6}$), and metals such as silicon[45] ($1.7 \times 10^{-6}$) and Fe[46] ($2 \times 10^{-7}$), where the values in parentheses are the abundances relative to that of hydrogen observed in the diffuse cloud ζ Oph. It is therefore not strange that most of the molecules detected, 70%, can be formed with just the four most abundant reactive elements, H, O, C, and N, while the remaining 30% make use of less abundant elements such as the second-row elements Si, S, and P, the halogens F and Cl, and metals. Apart from the obvious influence of the cosmic abundances of elements, perhaps the most striking fact is the prevalence of organic molecules, noticeably among the most complex ones. Around three-fourths of the detected molecules and all those with more than five atoms contain at least one carbon atom, while oxygen, which is more abundant than carbon, is a constituent of just one-third of the detected molecules. Therefore, interstellar chemistry is mostly a carbon-based chemistry, as is the chemistry of life, something which is certainly due to the exceptional chemical properties of carbon and its ability to form strong C–C bonds, allowing a large variety of complex molecules with a carbon skeleton to be formed. It is worth noting that while some of the interstellar molecules are well-known terrestrial species, such as $H_2O$, $NH_3$, CO, and $CO_2$, some others are very reactive and can hardly survive in terrestrial laboratories a sufficient time to allow for their characterization. In fact, some of them (e.g., HCO , HNC, $C_3N$, $C_4H$, $C_8H$, and $C_5N^-$) were discovered in space prior to their spectroscopic characterization in the laboratory.[49–57] It is also interesting to note that most of the detected molecules are electrically neutral, while only a few have charge. Molecular ions are scarce mainly because of their highly reactive character. In general, the degree of ionization inside molecular clouds is low, with most of the positive charge in the form of molecular cations, such as $HCO^+$, $N_2H^+$, $N_2D^+$, and deuterated $H_3^+$, while the negative charge is contained in dust grains and free electrons.[58] Nonetheless, the recent discovery of various molecular anions in interstellar and circumstellar media indicates that a substantial fraction of the negative charge is also in the form of molecular anions.[57,59–64] The molecule with the highest number of atoms detected has been for some years the cyanopolyyne $HC_{11}N$,[65] although evidence on the existence of polycyclic aromatic hydrocarbons (PAHs) with more than 50 atoms has existed for a long time.[66] The unequivocal identification of a PAH has however not been obtained yet, although the recent discovery of the fullerenes $C_{60}$ and $C_{70}$ in the ejecta of a dying star[67] indicates that indeed large carbon- based molecules are likely to be common in interstellar space.

Since its beginnings in the early 1970s, the study of interstellar chemistry has proven very fruitful in explaining how the chemical composition of interstellar clouds varies as matter is recycled along the life cycle of stars. It has also promoted very interesting collaborations among chemists and astronomers and constitutes nowadays a very active area of research, as demonstrated by the variety of topics covered within this current issue. Here we review the state of the art of chemical kinetics databases and networks, which contain rate constants of chemical reactions, information which is without a doubt the most essential input needed to build up any kind of chemical model. Enormous progress has been achieved in this area thanks to laboratory measurements, especially those carried out at very low temperature, and to theoretical calculations for both gas-phase and surface reactions. A particular emphasis is given to the successes and failures of chemical models of cold dark clouds, a type of interstellar region for which interstellar chemistry was originally devoted and which has often served as a benchmark to test chemical models.

## 2. BUILDING UP A CHEMICAL MODEL OF AN INTERSTELLAR CLOUD

The chemistry of the interstellar medium (ISM) is governed by gas-phase processes and interactions with interstellar grain surfaces. The most complete current chemical models aim at taking into account all these processes at the same time to compute the chemical composition of the gas and the ices at the surface of the grains. Some of the models adopt a simple geometry, assuming only one homogeneous shell of material (0D) with physical conditions not evolving with time. Others take into account a gradient of temperature and density increasing toward the center of the cloud and in some cases consider physical conditions evolving with time. Table 2 summarizes the level of complexity that can be included in chemical models. One-dimensional geometry and evolving physical conditions are briefly discussed in this section. In section 3.2, where we describe the formation of molecules in cold dense clouds, we focus on the most simple model: pure gas-phase chemistry with fixed homogeneous physical conditions (0D).

Table 2. Levels of Complexity of Chemical Models

| Chemistry | Pure gas-phase |
| --- | --- |
| | Gas-phase + adsorption and desorption |
| | Gas-phase + full treatment of gas-grain interactions and surface reactions |
| Physical conditions | Fixed |
| | Evolving with time |
| Geometry | 0D |
| | 1D |

### 2.1. Gas-Phase Processes

In the diffuse ISM, where the density of dust grains is small (dust densities below $10^{-9}$ grain/cm$^3$ and densities of H nuclei below $10^3$ cm$^{-3}$), photodissociations and ionizations by the ultraviolet (UV) radiation field, produced by surrounding massive stars, dominate the chemistry. Most species are thus in atomic form and even ionized for the atoms that have an ionization potential below 13.6 eV. Above this energy the photons are very rapidly absorbed by the ubiquitous atomic hydrogen. In denser regions (densities of H nuclei above $10^4$ cm$^{-3}$), UV photons are absorbed by the external layers of the clouds and only cosmic-ray particles can penetrate and produce ionization. In dense clouds, molecular hydrogen, the most abundant species, is ionized by cosmic-ray particles. The resulting high energetic electrons then interact again with the gas by exciting the $H_2$ molecules. The desexcitation of $H_2$ induces the emission of UV photons. This process, also known as the Prasad–Tarafdar mechanism,[68] is at the origin of a UV field inside dense clouds, which dissociates and ionizes species in the gas phase and at the surface of the grains. In chemical models, ionization and dissociation rates by cosmic-ray particles (direct and indirect processes) are given proportional to the total $H_2$ (or H) ionization rate, usually called $\zeta$ (s$^{-1}$).[69,70] Taking into account correctly the effect of cosmic rays in dense clouds is rather complicated since it depends on (i) the form of the cosmic-ray energetic distribution and (ii) its intensity and (iii) the physical conditions within the cloud.[71] For simplification, most chemical models assume a constant $\zeta$ in dense clouds, even when assuming a one-dimensional structure. Recently, Rimmer et al.[72] have found a dependence of this rate on the column density of gas in the cloud by studying gradients in the chemical abundances. More details on the effect of the cosmic rays on the chemistry can be found in a paper by Dalgarno.[73] Other types of reactions (bimolecular reactions) take place in the gas phase, such as ion–neutral reactions, neutral–neutral reactions, electronic dissociative recombinations, etc. Many reviews on these reactions exist in the literature, for instance, those by Canosa et al.,[74] Wakelam et al.,[75] Amith[76] and Larsson et al..[77]

Chemical models compute the composition of the gas (abundances of species) as a function of time, starting from an initial composition and for a set of parameters. The parameters of the models are the following:
- Temperature of the gas (the reaction rate coefficients depend in some cases on this parameter), T.
- Density of the gas (usually the density of H nuclei is used), $n_H$.
- Cosmic-ray ionization rate of $H_2$, $\zeta$.
- Elemental abundances (total abundances of helium, carbon, oxygen, etc.).
- Reaction rate coefficients.

Once those parameters have been defined, the model solves the following differential equations:

$$\frac{dn_i}{dt} = \sum_{j,k} k_{j,k} n_j n_k - n_i \sum_l k_{i,l} n_l, \qquad (3)$$

where $n_i$, $n_j$, $n_k$, and $n_l$ represent the densities of species i, j, k, and l, respectively, $k_{j,k}$ and $k_{i,l}$ represent the rate coefficients of the reactions between species j and k and between species i and l, respectively. Reactions {j,k} produce species i, whereas reactions {i,l} destroy it. These equations are written for all the species of a model, which are coupled by the reactions. These equations form a highly nonlinear system that needs numerical solvers such as ODEPACK.[78]

## 2.2. Grain Surface Processes

The importance of grain surface processes has been invoked for the first time for the formation of the most important species in the ISM: molecular hydrogen.[32] In a very basic way, due to their thermal agitation, species in the gas phase (they can be molecules or atoms) collide with interstellar grains. In some cases, these species can create van der Waals liaisons with the molecules at the surfaces (physisorption) and thus are depleted from the gas phase. The efficiency of this sticking is parametrized by a sticking probability between 0 and 1 (probably close to 1 at the low temperatures of dense clouds). Once on the surface, species can diffuse at the surface either by thermal hopping, jumping from one site to the other, or by a tunneling effect. Once two species meet, they can react. This is called the Langmuir–Hinshelwood mechanism.[75,79] Species can go back to the gas phase through several processes such as direct thermal evaporation,[36,80] cosmic-ray-induced evaporation,[81,82] exothermicity of surface reactions,[83] or photo-desorption.[84–87]

Chemical models that include the processes previously described, coupled to the gas phase, have to solve the following equations:

$$\frac{dn_i^s}{dt} = \sum_{m,n} k_{m,n}^s n_m^s n_n^s - n_i^s \sum_l k_{i,l}^s n_l^s - k_i^{des} n_i^s + k_i^{acc} n_i, \quad (4)$$

where $n^s$ are the densities of species at the surface of the grains. $K_{m,n}^s$ and $k_{i,l}^s$ are the rate coefficients of the reactions at the surface of the grains between species m and n and between species i and l, respectively. $k_i^{des}$ and $k_i^{acc}$ are the desorption and adsorption rates of species i. For a description of the rates and the parameters, we refer to publications by Hasegawa et al.,[36] Semenov et al.,[88] and Wakelam et al.[89] The formalism shown in eq 4 is known as the "rate equation" method, in which the efficiencies of the surface reactions are expressed similarly to the gas-phase rate coefficients. For other formalisms, such as stochastic methods and master equations, we refer to Wakelam et al.[75] for a review on the different approaches.

## 2.3. Databases and Chemical Networks

Databases can be uncritical, specific to a certain type of astrophysical environment, or limited to a given range of temperature. The NIST database,[90] for instance, is an extensive uncritical database of neutral–neutral reaction rate constants. Vincent G. Anicich did a compilation of ion–neutral reaction rate coefficients from the literature, which only exists in the format of a PDF file.[91,92] For particular applications such as the chemistry of the interstellar medium or planetary atmospheres, the KInetic Database for Astrochemistry (KIDA)[93] presents a compilation of uncritical data with recommendations over certain ranges of temperature for selected reactions. Those generic databases have to be as complete as possible; i.e., they must contain a large number of reactions and detailed information about the reactions such as the methods used to obtain the rate coefficients, the range of temperature, and the bibliographic information. The development and update of such databases is critical to improve the kinetic data used in the astrochemical models.

Any chemical model constructed for the interstellar medium relies on a list of processes and associated parameters that we will call "chemical networks". Chemical networks are created and updated from generic databases. There are two main chemical networks publicly available for interstellar matter applications: the UMIST Database for Astrochemistry chemical network[69,94,95] and the OSU chemical network.[96] Both networks have been developed for more than 20 years now. They are regularly updated according to the new data available. The most recent version of the OSU network is called kida.uva.2011.[70,93]

2.4. Reduced Networks

For some applications, reduction of the number of reactions is often required. Some physical models, for the collapse of clouds and the formation of stars, for instance, take into account the cooling and heating through the molecular and atomic lines. Some of these models also include the effect of the magnetic field and require information on the ionization fraction of the gas. These physical models are usually very time-consuming, and the coupling with the computation of the chemical abundances makes it impossible for the time being to use very large chemical networks. Several methods exist to reduce the number of reactions and species. The easiest and imprecise method is to ignore some families of chemical species and limit the complexity of the molecules. Nelson and Langer[97] and Oppenheimer and Dalgarno,[98] for instance, studied the formation of carbon monoxide in evolving isolated low-mass clouds and the ionization fraction in dense interstellar clouds, respectively. The chemical network was simply composed of 10 reactions, including some net reactions, to be combined with 3D hydrodynamical simulations. The quantitative aspect of the predictions obtained with those networks should however be taken with caution. As an example, Hincelin et al.[99] showed that the abundance of molecular oxygen is very sensitive to the rate coefficient of a reaction that does not involve oxygen: the $N + CN \rightarrow N_2 + C$ neutral–neutral reaction. Changing the rate coefficient of this reaction changes the abundance of CN, which can be one of the main consumers of O depending on its abundance.

Starting from larger networks and using network reduction techniques, Ruffle et al.[100] published a number of minimum chemical networks to compute the abundance of gas-phase CO in translucent regions. These networks contain tens of species and between 100 and 250 reactions depending on the physical conditions. Those reduction methods entail looking at the sensitivity of the computed abundances to the species and reactions. Below a certain criterion, those species and reactions are removed from the network.[101,102] The weakness of these methods is that the list of selected reactions depends on the initial large network. Any modification of this network (addition of new reactions or modification of rate coefficients) may change the reduced network. They are also only valid over a certain range of physical parameters for which they have been tested. Grassi et al.[103] proposed a new technique for network reduction, which has the advantage of being integrated into the model calculation itself. At each time step, fluxes of production and destruction of species are tested without slowing the calculation. This new method is then independent of time, the physical conditions, or the initial network and does not necessitate any preliminary treatment.

2.5. Model Uncertainties and Limitations

2.5.1. Gas-Phase Processes. Among the sources of uncertainties of chemical models, three can be clearly identified:

(1) Uncertainties in the initial conditions.

(2) Incompleteness of the chemical networks.

(3) Uncertainties in the model parameters.

The time scale to reach steady state in the interstellar medium is longer than the dynamical time scale of the physical condition evolution in most regions; as a consequence, the chemical composition depends on the initial conditions (initial chemical composition). For example, under dense cloud conditions (typical temperature of 10 K, density of a few $10^4 \, \text{cm}^{-3}$, and a visual extinction $A_V$ of 30 magnitudes[104]), the typical time to reach the steady state for a reservoir molecule such as CO is approximately $10^9$ yr if both gas-phase chemistry and gas–grain interactions are considered, whereas the typical lifetime of such objects is $10^7$ yr or shorter.[105,106] Since most chemical models of dense clouds do not take into account the formation of the cloud itself, the computed chemical composition depends on the initial conditions.

The incompleteness of the chemical networks is probably the most difficult to quantify. It is usually the disagreement between the model predictions and the observations of molecules that reveals missing processes. One example is the nonthermal evaporation process of surface molecules due to the release of energy by exothermic surface reactions.[83] This process postulates that the energy released by such a chemical reaction is not completely lost to lateral translation along the grain surface but allows a fraction of the produced species to be evaporated in the gas phase, breaking the bonds to the surface. The addition of this process in chemical models was guided by the observation of gas-phase abundances of methanol ($CH_3OH$) in cold dense clouds that could not be reproduced by the models. For several years, it was believed that methanol could be formed in the gas phase by the radiative association of $CH_3^+$ and $H_2O$ followed by the dissociative recombination with electrons of the $CH_3OH_2^+$ ion. The rate constant of the radiative association was shown experimentally to be much smaller at low temperature than what was previously estimated[107] and the branching ratio of the dissociative recombination much smaller for the channel leading to methanol.[108] Methanol is produced very efficiently on the surface of the grains through the successive hydrogenation of CO. The inclusion of nonthermal evaporation due to exothermic reactions at the surface of the grains in the models allows some of the methanol produced on the surface to be released to the gas phase and contribute to the observed gas- phase abundances. The efficiency of this process remains however very uncertain, and only an upper limit has been derived from laboratory experiments on interstellar ice analogues.[109] As a consequence, the results are very parameter-dependent. The recent detection of complex organic molecules other than methanol in cold dense clouds, such as dimethyl ether and methyl formate,[110,111] indicates also that nonthermal desorption processes may indeed be at work in such cold and quiescent environments.

The uncertainties in the model parameters are the most easy to address. The parameters involved in a chemical model are on one side the rate coefficients of the chemical processes included in the chemical network and on the other the physical conditions adopted for the cloud: density, temperature, cosmic- ray ionization rate, etc. The propagation of the model parameter uncertainties during the calculations of the abundances by the model can be studied with Monte Carlo simulations.[112,113] In those simulations, the model is computed for different sets of parameters and the computed abundances are modified according to their sensitivity to the parameter. A large number of models have to be run to properly cover the range of uncertainties. Each parameter can be changed separately or together to study combined effects. Figure 1 shows an example of such simulations. The abundance of $HC_3N$ in the gas phase, computed by a pure gas-phase model, is shown for typical dense cloud conditions (a gas and dust temperature of 10 K, a density of H nuclei of $2 \times 10^4 \, cm^{-3}$, and a visual extinction of 30 magnitudes). In this case, only the cosmic-ray ionization rate has been randomly changed between $10^{-16}$ and $10^{-17} \, s^{-1}$. In the figure, the probability density function is shown to indicate the spread of the curves. At $10^5$ yr, 95% of the uncertainty in the predicted abundances is less than 1 order of magnitude. The impact of all the parameter uncertainties (temperature, density, cosmic-ray ionization rate, etc.) can be studied in this way.[113]

Figure 1. Density probability function of the predicted abundance (compared to that of H) of $HC_3N$ as a function of time for dense cloud conditions, computed by a gas-phase model in which the cosmic-ray ionization rate is varied between $10^{-16}$ and $10^{-17}\,s^{-1}$.

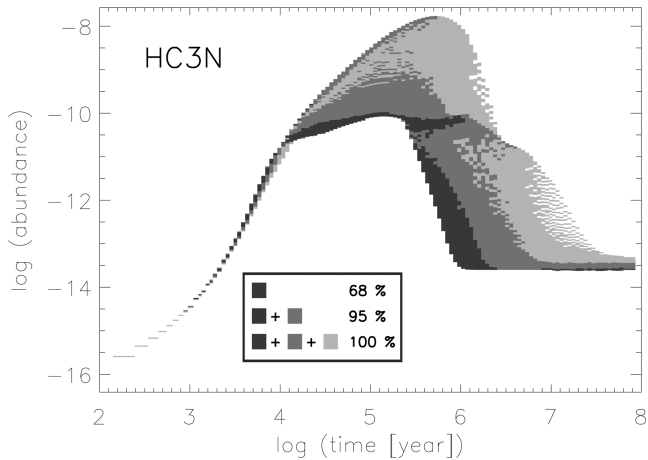

2.5.2. Grain Surface Processes.

Defining the uncertainties in the surface processes is more complex because of the intrinsic nature of the processes. In addition to the uncertainties in the parameters of the surface chemistry, such as the binding energies of the molecules and the number density of grains, uncertainties in the processes themselves exist. It is usually assumed, for example, that gas cations cannot stick to the surface of the grains because the grains are mainly negatively charged. The negative charge of the grains may however not be completely delocalized, and cations could be at least partly present on the surfaces. The possible identification of $NH_4^+$ and $OCN^-$ in interstellar ices could be an indication of such presence.[114] Gas-phase species could also produce real chemical liaisons with the species of the bare grains (chemisorption),[115] considerably reducing the mobility of these species on the surface. Reactions on the surfaces could also be produced by hot atoms coming from the gas phase that directly heat a species on the surface (Eley–Rideal mechanism[116]). Another strong source of uncertainty is the nature of the chemical composition of the grains, their roughness, size, and shape. The nature of the surfaces, their roughness and their shape, defines the strength of the bonds between adsorbed species and the surface and the mobility of the species on the surfaces.[117] The size of the grains defines the surface available for the depletion of the gas-phase species. The last tricky aspect of the surface chemistry is the fact that gas-phase molecules are absorbed on grains in layers. In addition to the fact that the binding energies may not be the same for all the layers (the binding energy of CO on top of silicates is not the same as that of CO on top of water ice), it is not obvious if the molecules can diffuse (and react) between different layers.

Finally, the presence of small PAHs (polycyclic aromatic hydrocarbons) can also influence the chemistry at several levels. In the diffuse regions, the atomic hydrogen from the gas phase can be adsorbed onto PAHs by chemisorption and form molecular hydrogen in these hot and irradiated regions.[118] In denser regions, PAHs are mostly negatively charged so that if their abundance is larger than that of electrons (a few $10^{-8}$ with respect to that of $H_2$), electronic recombination with PAH$^-$ may exceed that with free electrons.[119,120] The uncertainty in this matter comes from the fact that the size distribution of PAHs in dense regions is not known. In addition, there are almost no experimental or theoretical studies of the products of the electronic recombination of cations with PAH$^-$.

## 3. CHEMICAL MODELS OF COLD DENSE CLOUDS

Cold dense clouds have provided since the beginnings of astrochemistry one of the preferred astronomical laboratories on which to test chemical models. On one hand, the large variety of molecular species proven by radioastronomical observations in some remarkable dark clouds such as TMC-1, where more than 60 molecules have been observed to date (see Table 4), poses an important challenge for astrochemical models. On the other, the quiescent environment of dark clouds provides a relatively simple scenario, as compared with other astronomical regions, on which to apply chemical models. In the central regions of starless cores, the lack of an internal luminous source and the high extinction of external interstellar light provided by the dust content of the cloud result in a very low gas kinetic temperature (around 10 K). Under these conditions, high-temperature gas-phase chemistry and photo- chemistry induced by interstellar UV photons are suppressed, while desorption processes, which replenish the gas phase with material processed on the surface of dust grains, are minimized.

### 3.1. Examples of Published Works

If one is forced to choose a starting point in the history of chemical models of cold dense clouds, that starting point probably needs to be searched for in the studies published in the early 1970s by Watson[28–30] and Herbst and Klemperer.[31] These studies were able to identify the main processes that control the chemistry of cold dense clouds: (I) It was pointed out that hydrogen is mainly present as $H_2$ in the gas phase, although it must be formed on the surface of dust grains. (II) Cosmic rays were identified as the main ionizing agent in the interior of dark clouds, leading to the formation of the molecular ion $H_3^+$, which occupies a cornerstone position in interstellar chemistry. (III) The importance of gas-phase ion chemistry for the buildup of molecular complexity was also recognized. In particular, it was suggested that reactions of proton transfer from $H_3^+$ as well as H atom abstraction reactions of the type $AH_n^+ + H_2 \rightarrow AH_{n+1}^+ + H$ (where A stands for a heavy atom such as C, N, and O, and n = 0, 1, 2,...) would allow formation of simple hydrides

$AH_n^+$ with one heavy atom, while ion–neutral reactions involving two heavy species would do the work to synthesize species with more than one heavy atom (e.g., $C^+ + CH \rightarrow C_2^+ + H$). These studies also pointed out that positive molecular ions would be largely depleted by dissociative recombination with electrons, leading to the formation of neutral fragments. It was also suggested that some ion–neutral radiative associations could be important in the synthesis of molecules if other reaction channels are closed.[31]

A large number of chemical models of cold dense clouds have been developed since the early 1970s to the present day. The situation prior to 2000 has been reviewed from various points of view.[121–125] The study of Herbst and Klemperer[31] was the first to carry out a chemical model of a cold dense cloud by solving the coupled equations of a chemical kinetics system of 37 species and 100 reactions under steady-state conditions and yielding quantitative predictions of molecular abundances. Later on, Mitchell et al.[126] pursued this same line of thought and carried out an extensive study of the chemistry of dark clouds by using a list with about 400 gas-phase reactions involving 100 different species.

An alternative explanation for the formation of molecules in cold dense clouds was proposed by Allen and Robinson[127] based on previous ideas on $H_2$ formation:[35] recombination of neutral fragments on the surface of dust grains and further molecule ejection driven by the chemical heat supplied upon bond formation. The model of Herbst and Klemperer,[31] which based the formation of molecules on gas-phase chemical reactions, and the model of Allen and Robinson,[127] where molecular synthesis relies exclusively on dust surface reactions, represent two drastically different approaches, although both yielded calculated abundances that were in good agreement with those derived from observations at that epoch. The idea of molecule formation taking place on grains was not pursued by subsequent chemical studies of cold dense clouds, mainly because it was not clear how species much heavier than $H_2$ would be desorbed from grain surfaces in the cold and quiescent conditions of dark clouds. Nowadays, however, the role of grain surface reactions and nonthermal desorption processes in cold dense clouds is coming back to scene due to the observation of some molecules such as $CH_3OH$ (see below), whose formation is difficult to explain by gas-phase chemistry.

It was therefore the line of thought of molecule formation in the gas phase that was pursued by subsequent chemical models of dark clouds. An important upgrade to such models was to include the time dependence. Usually the evolution in the abundances of the different species was calculated from an initial set of abundances under fixed physical conditions (temperature, density, and cosmic-ray ionization rate). This approach was later named the "pseudo-time-dependent chemical model" by Leung et al.[128] to distinguish it from a full-time-dependent model, in which also the physical conditions are allowed to evolve with time. The first efforts to this end were undertaken in the late 1970s by Iglesias,[129] Suzuki,[130] and Prasad and Huntress.[131,132] These first pseudo-time-dependent chemical models of cold dense clouds indicated that chemical abundances reach the steady state after $\sim 10^7$ yr, a time which is comparable to or slightly higher than the lifetime of dark clouds, as inferred from dynamical considerations on the time scale of gravitational collapse of the cloud.[105,106] The fact that dark clouds may not have attained a steady state in their chemical composition within their lifetimes stressed the need to study the time dependence of molecular abundances.

Since the late 1970s, a large number of pseudo-time- dependent chemical models of dark clouds have investigated various aspects of cold dense cloud chemistry and, ultimately, have tried to address the question of whether gas-phase processes can account for the molecular abundances observed in dark clouds. The sensitivity of the chemistry to the elemental abundances was investigated by Graedel et al.,[133] who introduced the concept of "low-metal" elemental abundances, in which the abundances of metals, silicon, and sulfur are depleted by a factor of 100 with respect to the values typically used until then (those observed in the diffuse cloud ζ Oph with the Copernicus satellite[134]). This choice of elemental abundances resulted in a better agreement with observations, in particular with the ionization degree in dark clouds as inferred from observations of $HCO^+$,[135] and has therefore been adopted by many subsequent chemical models of cold dense clouds. Various studies have also addressed the question of how and to what extent gas-phase chemistry synthesizes moderately complex molecules.[128,136–138] Leung et al.[128] made a first effort to include a significant number of complex molecules in their pseudo-time-dependent model and found that their formation is favored if the abundance of neutral atomic carbon is high and the C/CO abundance ratio is close to unity, conditions that occur at a so-called "early time" ($10^5$–$10^6$ yr), well before the time needed to reach steady-state abundances ($\geq 10^7$ yr). Subsequent gas-phase pseudo-time-dependent models[138–140] have often found that the abundances calculated at or near the early time are overall in good agreement with the abundances observed in the dark cloud TMC-1 (often used as a reference to test the successes and failures of dark cloud chemical models). This fact validates somehow the gas-phase scheme for the formation of molecules, although much debate exists on the physical meaning of such early time.[141,142]

The chemistry of specific elements in cold dense clouds has also been investigated in detail. For example, the chemistry of silicon was studied by Herbst et al.,[143] who predicted SiO to lock up most gas-phase silicon in dark clouds, with an abundance well above the upper limits derived from observations. In the case of phosphorus, Millar[144] calculated an abundance for PN again above the observational upper limits. If true, these models point toward very high depletion factors for Si and P (>$10^6$ and $\sim 10^4$, respectively) in cold dense clouds, although it has also been suggested that no extreme depletion needs to be invoked for silicon if the major Si- containing species is $SiH_4$ rather than SiO.[145] The situation has not changed much since then, as no Si- or P-bearing molecule has ever been observed in cold dense clouds.

Some interesting chemical effects that become apparent at very low temperatures, such as isotopic fractionation and ortho/para chemistry, are best proven in cold dense clouds and have also been studied through chemical models. Isotopic fractionation consists in the enrichment in heavy isotopes that a given molecule may experience, the origin of which is the fact that a heavy isotopologue is thermodynamically, and kinetically if there are suitable fractionation reactions, favored over a lighter one due to the lower zero-point energy of the heavier species. Isotopic fractionation has been addressed by chemical models of cold dense clouds for various elements,[146−152] although the most extreme effects occur in the case of deuterium. Up to a triply deuterated molecule, such as $ND_3$, has been observed in a dark cloud with an $ND_3$/$NH_3$ abundance ratio of ~8 × 10$^{-4}$, which implies an enhancement of more than 10 orders of magnitude over the purely statistical value expected from the abundance of deuterium in the interstellar medium.[153] Along the same line, there are interesting chemical effects related to the para and ortho forms of molecular hydrogen. Some chemical models of cold dense clouds have studied the influence of the different reactivities of para and ortho states of $H_2$ on aspects such as the formation of ammonia,[154] the ortho-to-para ratio of cyclic $C_3H_2$,[155] or its utility to put a constraint on the age of dark and prestellar cores.[156,157]

In 1983, Prasad and Tarafdar[68] realized that the impact of energetic secondary electrons produced by cosmic-ray ioniza- tion can promote $H_2$ to excited electronic states from which spontaneous emission would produce UV photons with a sufficient energy to dissociate many molecules. The effects of this ultraviolet radiation inside dark clouds on the chemistry was subsequently investigated by Sternberg et al.[158] and Gredel et al.,[159] who found that there were no dramatic consequences, although large hydrocarbons would lower their abundances due to photodestruction. There is observational and experimental evidence that these UV photons could significantly modify the composition of the ices at the surface of the grains[160] and provide an efficient mechanism to desorb molecules from grain mantles, although its extent may be limited to low-density (<10$^3$ cm$^{-3}$) regions.[161]

3.1.1. Improving the Kinetic Data.

Much effort of chemical models of cold dense clouds has been put into the upgrade of chemical networks. Rate constants have been continuously re-evaluated according to new measurements and state of the art theoretical calculations, while many have been estimated for new chemical processes involving increasingly larger molecules. A good number of studies have evaluated the impact of modifications in the reaction schemes on the calculated abundances. Herbst and Leung[162] investigated the effect of an increase in the rate constants of reactions between ions and polar neutral species, following the previous idea that such reactions occur faster at the low temperatures of dark clouds than at room temperature,[163] and found that the calculated abundances of polar neutral molecules are generally reduced using these enhanced rate coefficients. In the process of extending chemical networks, however, important discrep- ancies among different authors were not rare. For example, the pseudo-time-dependent model of Millar and Nejad[136] found substantial differences in the calculated abundances of complex molecules with respect to the model of Herbst and Leung,[128,137] with a better agreement with observed abundances in TMC-1 found at late times ($\sim 10^7$ yr) rather than at early times ($\leq 10^6$ yr), as opposed to the findings of Herbst and Leung. A subsequent study by Millar et al.[164] was able to identify the sources of such discrepancies as arising from different assumptions on reactions involving O atoms and branching ratios of dissociative recombinations of complex ions. This latter issue was addressed by Millar et al.,[165] who estimated the branching ratios of a good number of dissociative recombinations between polyatomic ions and electrons according to Bates idea[166] that the preferred channel is the one involving the least rearrangement of valence bonds. Later on, it was, however, demonstrated by experiments[77,167] that dissociative recombinations do not tend to follow the simple Bates rule and that there is significantly more fragmentation than previously thought, although calculated abundances for cold dense clouds have been quite surprisingly found to be not very sensitive to these new findings.[168]

Another major upgrade to chemical networks for cold dense clouds have come from the revision of rate constants for neutral–neutral reactions. Historically, reaction schemes constructed to study the chemistry of dark clouds have mainly relied on ion–neutral reactions, the number of neutral–neutral processes being limited to a few exothermic reactions between radicals, which were thought to occur without activation energy. Several experiments carried out since the 1990s, especially those using the CRESU technique, have demonstrated that a variety of exothermic neutral–neutral reactions, between two radicals and more surprisingly between a radical and one stable molecule, are barrierless and therefore quite rapid at the low temperatures of cold interstellar clouds. Various studies have addressed the effects that such rapid neutral–neutral reactions have on the output results of chemical models of dark clouds.[40,140,169] In general, it has been found that the abundances of complex molecules are reduced due to reactions involving O atoms, which tend to hamper the buildup of large hydrocarbons and cyanopolyynes.

The different updates of chemical networks devoted to the study of cold dense clouds have often led to a decrease in the abundances of complex molecules and to a worse agreement with the abundances observed in the dark cloud TMC-1. This may suggest that historical reaction schemes have been to some extent too optimistic in forming complex molecules and somewhat biased to successfully reproduce the abundances observed in chemically rich cold dense clouds such as TMC-1. Efforts to quantify the uncertainty that reaction networks produce in the abundances calculated by chemical models of dark clouds have been undertaken in various stud- ies.[112,113,170,171] These studies served also to identify important reactions so that an emphasis can be put on a better understanding of their rate constants to make more secure the predictions of chemical models.

### 3.1.2. Gas–Grain Models.

Many of the published chemical models of cold dense clouds have relied exclusively on gas- phase processes (except for the formation of $H_2$ on grain surfaces), aiming at evaluating the ability of gas-phase chemistry to account for the molecular abundances observed in dark clouds. Various studies, however, have worried about the consequences that the interaction between gas and dust grains may have for the chemistry of dark clouds.[36,82,172–182] The most immediate consequence is that, in the absence of efficient desorption processes able to return condensed material to the gas phase, all gas-phase species would be depleted onto grains on a time scale of $[(2-3) \times 10^9]/n_H$ yr (where $n_H$ is the density of H nuclei (cm$^{-3}$)).[127] Pseudo-time-dependent chemical models where desorption processes are not very important have shown that gas-phase abundances are not severely affected at a so-called "early time", although they are severely depleted at late times. In turn, grain surface reactions tend to form hydrogenated molecules on grain mantles, which become dominated by water ice.[36,173] Various desorption mechanisms have been proposed to counterbalance accretion in cold dense clouds, among them transient heating and explosions induced by the impact of cosmic-ray heavy Fe nuclei,[81] and propagation of magneto-hydrodynamic waves within the cloud,[183,184] photodesorption by cosmic-ray-induced UV photons,[161] mantle explosions induced by exothermic radical recombination reactions,[172] and ejection via the energy released by exothermic chemical reactions.[83,127] Thermal evaporation and photo- desorption by interstellar UV photons are not effective in the interior of dark clouds due to the low temperatures and the opaque character of the cloud.[185] Willacy and Millar[178] showed that the UV photons induced by cosmic rays inside the clouds were not sufficient as well to significantly produce evaporation. Direct cosmic-ray heating and surface reactions seem to be efficient and complementary to produce partial or total mantle desorption, although some uncertainties in their efficiency are still a matter of debate. Shen et al.,[186] for instance, concluded that the formalism proposed by Hasegawa and Herbst[82] for direct cosmic-ray desorption overestimated the efficiency of the process and that mantle explosions induced by exothermic radical recombination reactions are the dominant nonthermal evaporation process in dark clouds. Kalvāns and Shmeld,[160] on the contrary, suggest that the porosity and layering nature of the ices prevent this process from being efficient. The calculated chemical composition of grain mantles being quite different from that of the gas phase, the efficiency of desorption processes becomes critical to determine the extent at which the gas-phase composition is modified by grain surface chemical processing. Unfortunately, grain surface chemical reactions and nonthermal desorption processes are not well understood, which hampers the ability of gas–grain models to provide secure predictions. The situation will hopefully improve in the next few years since surface chemistry has become a very active area of research (see the reviews by Garrod and Widicus Weaver,[37] Watanabe,[38] and Cuppen[39] in this issue).

### 3.1.3. Bistability.

In the early 1990s, an interesting phenomenon related to chemical models of interstellar clouds was discovered. Le Bourlot et al.[187] realized that there are two possible solutions to steady-state chemical models of dark clouds, called low- and high-ionization phases (LIP and HIP, respectively), the ionization degree and C/CO abundance ratio being higher in the latter. Such double solution was also found by pseudo-time-dependent models starting from different initial abundances (same elemental abundances but elements put into different species). The bistable behavior of kinetic rate equations, with two physically meaningful steady-state solutions, arises mathematically from the high number of nonlinearities in the equations and, interestingly, implies that both solutions could coexist in dark clouds depending on the initial chemical composition. Bistability has been found to occur within a certain range of densities around $10^4$ cm$^{-3}$, its exact location being sensitive to parameters such as the cosmic- ray ionization rate, the gas-phase abundance of sulfur, and the C/O abundance ratio.[187,188] It has also been found that the extent to which bistability occurs is restricted or even suppressed if interactions between gas and dust and grain surface reactions are considered in the kinetic equations[189–191] and that its extent is also quite dependent on the reaction network used.[171,192] Despite a few attempts to find observational evidence of this bistability in the interstellar medium,[193] the more we complete the chemical networks by adding missing reactions, the more the domain of physical parameters where bistability can be found shrinks.[171]

3.1.4. Geometry and Time Dependence of the Models.

The chemical models of dark clouds described so far are of the "one-point" variety, in which the chemical composition of a cell of gas and dust with static physical conditions is calculated either at the steady state or as a function of time from some initial abundances. Dark clouds are, however, not homogeneous but have density and temperature gradients, and molecular abundances may vary by several orders of magnitude at subparsec scales, as probed by mapping observations.[194,195] If one aims at studying the spatial structure of the molecular abundances, then the geometry of the source needs to be taken into account. There have been two major approaches to this problem. The first entails simultaneously solving the thermal balance and chemical equations at the steady state.[196] This approach has been widely used to study the so-called PDRs (photon-dominated regions), which are interstellar clouds illuminated by an ultraviolet field, and more recently it has also been applied to protoplanetary disks.[197,198] In the case of cold dense clouds, the interior is usually well protected from external UV photons but the outer layers can be considered as a subclass of PDR, the study of which is beyond the scope of this paper (see the review by Hollenbach and Tielens[199]). In the second approach a physical structure is adopted for the dark cloud, either calculated from hydro-dynamical considerations or obtained from observational constraints of a given source, and a series of independent pseudo-time-dependent chemical models are run at the local physical conditions of different positions across the cloud.[200,201] This latter approach has been adopted by Nejad and Wagenblast[202] to study the time-dependent chemistry as a function of the depth into a spherical cold dense cloud from the inner UV-shielded regions, where molecular synthesis is favored, to the outer regions, where most molecules are photodissociated.

Probably the most ambitious approach undertaken to model the chemistry of cold dense clouds is a model in which the time dependence of chemical abundances is calculated taking into account the evolution of the physical conditions of the cloud, i.e., what we call a "full-time-dependent chemical model". In pseudo-time-dependent chemical models of dark clouds, the chemical time scales (the abundances of complex molecules peaking at a so-called "early time" of a few $10^5$ yr and the steady state reached at about $10^7$ yr) are of the same order of magnitude as the estimated cloud lifetimes ($10^6$–$10^7$ yr),[105,106] which implies that chemical processes may well occur on a time scale similar to that of the collapse of the cloud. Moreover, other intermittent dynamical processes such as shock waves or turbulent motions are likely to occur on time scales shorter than those of the chemical processes. A pseudo-time-dependent chemical model, which considers static physical conditions and usually adopts initial abundances typical of diffuse clouds, may provide a realistic view of the chemistry of dark clouds if a rapid collapse occurs from the precursor diffuse state. This scenario, however, is likely to be too idealized in view of the above arguments, which has motivated the development of full-time- dependent chemical models. On one hand, such models allow some of the oversimplifications of pseudo-time-dependent models to be overcome. These are mainly the fact that physical conditions remain static during the chemical evolution, but also the problematic issue of the initial abundances. Note that as far as dark clouds may probably not have attained a chemical steady state, transient abundances become more relevant, and these depend to a large extent on the choice of the initial abundances. Of course, in the frame of a full-time-dependent model where chemistry and physics evolve from a diffuse state, it seems well justified to adopt initial abundances typical of diffuse clouds. On the other hand, the details on the process of cloud collapse are not well understood (several scenarios are plausible), and this adds an important uncertainty to the salient results of the full-time-dependent model. A common scenario adopted to face up to the problem has been to consider the hydrodynamic collapse of a cloud from a diffuse state until the formation of a dense core. Such an approach was adopted in the pioneering study by Gerola and Glassgold[203] and the subsequent more extensive work by Tarafdar et al.[204] This latter study found, among some other interesting conclusions, that clouds spend most of their lives in a diffuse state ($A_V \leq 2$) and that once the core becomes dense ($10^3$–$10^4$ cm$^{-3}$) its collapse occurs extremely fast. The chemistry used by Tarafdar et al. is rather simple, but under such a scenario, it is likely that during the long-lived diffuse state the low density and high exposure to UV light prevent an efficient synthesis of complex molecules, which would be formed during the second short-lived dense stage. A fast collapse from a precursor diffuse state seems to support the use of pseudo-time-dependent models, although it should be noted that the model of Tarafdar et al. neglects magnetic fields and rotation, which are known to retard the collapse of the cloud.[205]

The question of why in pseudo-time-dependent chemical models early-time abundances are in better accord with observed values than abundances calculated at late times (especially in the case of complex species such as carbon chain molecules when compared to sources such as TMC-1) has been addressed by various authors. Williams and Hartquist[142,206] have argued that interstellar clouds would be subjected to dynamical cycles between a dense phase (formed by gravita- tional collapse) and a diffuse state (driven by stellar winds which would dissipate dense clumps). This idea has been explored in the context of dynamical and chemical models.[207–209] In the same vein, Prasad et al.[17] have suggested a scenario in which dense cores may evolve toward a star- forming region only under certain conditions; otherwise, they may revert to a diffuse state. Since the time spent by a dynamically evolving cloud in a given phase decreases as the density of the core regions increases, chemistry would just have a limited time to proceed during a dense phase. An alternative scenario has been proposed by Chieze et al.,[141] in which the mixing of material between diffuse and denser regions of interstellar clouds would allow preservation of the calculated early-time abundances for much longer periods of time.

3.2. Molecule Formation in Cold Dense Clouds: State of the Art

The extremely rarefied and cold gas of dark clouds seems not the best environment for the buildup of molecules. Nonethe- less, radioastronomical observations have shown that cold dense clouds contain a large wealth of exotic and complex molecules. More than 40 yr of research by astronomers and chemists have led to a good overall understanding of the chemical processes which are responsible for the in situ formation of molecules in dark clouds, although there are still some chemical issues which are barely or not at all understood. The terms "cold dense clouds" and "dark clouds" already contain their three main characteristics. The moderate densities, low temperatures, and high opacities are key for the chemistry that can take place in such an environment, which of course is completely out of thermochemical equilibrium and therefore determined by chemical kinetics.[210] The low temperatures of cold dense clouds imply that only reactions which are exothermic and have no activation barrier can play a role. In spite of the adjective "dense", densities are moderately low so that three-body reactions are of no relevance and chemical transformations occur on long time scales ($>10^5$ yr). In turn, the opaque character of the cloud avoids the penetration of ultraviolet radiation from the ambient interstellar field, so the only photochemistry that takes place in the internal regions is due to the relatively weak UV field caused by the impact of cosmic rays with $H_2$ molecules through the Prasad– Tarafdar mechanism.[68]

The chemistry in dark clouds is initiated by the cosmic-ray ionization of $H_2$ and He. Since helium has a large ionization potential (24.6 eV), $He^+$ ionizes easily most neutral species other than $H_2$ (the reaction between $He^+$ and $H_2$ is highly exothermic but has a large activation barrier). In turn, $H_2^+$ reacts rapidly with $H_2$ to form $H_3^+$. Molecular hydrogen having a proton affinity lower than those of most atoms and molecules, $H_3^+$ reacts rapidly with most neutral species through proton transfer:

$$A + H_3^+ \rightarrow AH^+ + H_2 \quad (5)$$

where A stands for any neutral species. Therefore, the ionization of most species caused by $He^+$ and their protonation by $H_3^+$ open the pathways to molecule formation in cold dense clouds. Positive ions may be hydrogenated by successive reactions with $H_2$:

$$A^+ + H_2 \rightarrow AH^+ + H \quad (6)$$

where $A^+$ is an arbitrary positive ion. The ionic pathways to form neutral molecules usually involve, as the last step, the dissociative recombination of the molecular positive ion with an electron, a process that may be schematically described as

$$AB^+ + e^- \rightarrow A + B \quad (7)$$

where $AB^+$ represents any generic molecular ion. These reactions allow formation of simple hydrides such as water through the sequence of reactions

$$O \xrightarrow{H_3^+} OH^+ \xrightarrow{H_2} H_2O^+ \xrightarrow{H_2} H_3O^+ \xrightarrow{e^-} H_2O$$

In the case of hydrides such as methane, the sequence of reactions of the type described in eq 6 is interrupted at some point by some reactions which do not proceed, in which case a reaction of radiative association between a positive ion and $H_2$ of the type

$$A^+ + H_2 \rightarrow AH_2^+ + h\nu \quad (8)$$

may aid to fill the gap. For example, the synthesis of methane in dark clouds occurs preferentially through the sequences of reactions

$$C^+ \xrightarrow{H_2} CH_2^+$$

and

$$C \xrightarrow{H_3^+} CH^+ \xrightarrow{H_2} CH_2^+$$

followed by

$$CH_2^+ \xrightarrow{H_2} CH_3^+ \xrightarrow{H_2} CH_5^+ \xrightarrow{e^-} CH_4.$$

which involve the radiative associations with $H_2$ of the $C^+$ and $CH_3^+$ ions. The ionic pathways to form large molecules containing more than one heavy atom (e.g., C, N, and

O) involve reactions of insertion of the heavy atom (either in the neutral state or ionized), such as

$$A^+ + A_nH_m \rightarrow A_{n+1}H^+_{m-1} + H, \quad (9)$$
$$\rightarrow A_{n+1}H^+_{m-2} + H_2,$$

$$A + A_nH^+_m \rightarrow A_{n+1}H^+_{m-1} + H, \quad (10)$$
$$\rightarrow A_{n+1}H^+_{m-2} + H_2,$$

where A represents a heavy atom, as well as reactions between two heavy fragments, either "condensation" reactions such as

$$A_xH^+_y + A_nH_m \rightarrow A_{x+n}H^+_{y+m-1} + H, \quad (11)$$
$$\rightarrow A_{x+n}H^+_{y+m-2} + H_2,$$

or radiative associations, which may be described as

$$A^+ + B \rightarrow AB^+ + h\nu, \quad (12)$$

where both $A^+$ and B fragments contain at least one heavy atom. We note that ion–neutral reactions may indeed allow formation of increasingly larger molecules, although they may also result in the destruction of large molecules if reaction channels leading to an important degree of fragmentation dominate. Other channels of hydrogen abstraction can also exist. Charge transfer reactions play a key role in the redistribution of the charge among the species with the lowest ionization potentials. They may be schematically described as

$$A^+ + B \rightarrow A + B^+, \quad (13)$$

where A and B are generic species such that A has a larger ionization potential than B so that the reaction is exothermic. Ionic pathways account for much of the molecular synthesis in cold dense clouds (see a recent review by Larsson et al.[77]), although neutral–neutral reactions in which at least one of the reactants is a radical may also play an important role. Among them, insertion reactions of the type

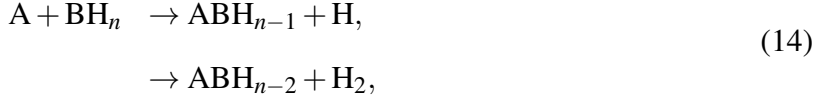

$$A + BH_n \rightarrow ABH_{n-1} + H, \quad (14)$$
$$\rightarrow ABH_{n-2} + H_2,$$

where A is a radical (either a heavy atom or small neutral fragment) and $BH_n$ any arbitrary H-containing molecule, can actively participate in the growth of molecules.

Apart from gas-phase chemistry, processes related to dust grains may also have a considerable effect on the chemical composition of dark clouds, although their role is much less clear. First, accretion of gas species onto dust grains is clearly an important process at late times. Second, grain surface reactions are likely to determine the chemical composition of grain mantles and may also influence to some extent the gas-phase chemical composition depending on the efficiency of non- thermal desorption processes.

How good is our knowledge of the chemical processes at work in cold dense clouds? What are the successes and failures of state of the art chemical models? A useful exercise to address these questions is to build a pseudo-time-dependent chemical model using state of the art chemical networks and compare the salient results with observed abundances in dark clouds. Our aim is to concentrate on purely chemical issues. In our one- point pseudo-time-dependent model, we have adopted a fixed temperature of 10 K and a fixed density of hydrogen nuclei of $2 \times 10^4$ cm$^{-3}$, typical values for cold dense clouds usually adopted in many previous studies. We consider that our point has a visual extinction of $A_V = 30$ to the edge of the cloud. This rather arbitrary value is chosen to suppress any photochemical processes due to external interstellar UV photons, since we are mainly interested in modeling the interior of a dark cloud. The visual extinction from the center to the edge of chemically rich dark clouds such as TMC-1 and L134N, as inferred from observations, is larger than 5,[195,211,212] and at this value photochemical effects are already very limited.[137] The cosmic- ray ionization rate of $H_2$ is set at $1.3 \times 10^{-17}$ s$^{-1}$, a value commonly used in many previous studies and which is at the lower limit of the range of values inferred for dense clouds $((1-5) \times 10^{-17}$ s$^{-1})$.[73]

Table 3: Initial elemental abundances

|  | Abundance relative to H |  |  |
|---|---|---|---|
| He | $8.5 \times 10^{-2}$ [a] | P$^+$ | $3 \times 10^{-9}$ [d] |
| C$^+$ | $1.8 \times 10^{-4}$ [b] | Na$^+$ | $2 \times 10^{-9}$ [d] |
| O | $3.3 \times 10^{-4}$ [b] | Mg$^+$ | $7 \times 10^{-9}$ [d] |
|  | $1.3 \times 10^{-4}$ [c] | Fe$^+$ | $3 \times 10^{-9}$ [d] |
| N | $6.2 \times 10^{-5}$ [b] | Cl$^+$ | $4 \times 10^{-9}$ [d] |
| S$^+$ | $8 \times 10^{-8}$ [d] | F | $1 \times 10^{-8}$ [e] |
| Si$^+$ | $8 \times 10^{-9}$ [d] |  |  |

[a] Solar value (Asplund et al.[213]). [b] ζ Oph value (C/O=0.55 ; Jenkins[214]). [c] Low oxygen abundance (C/O =

1.4; extrapolation to $n_H = 2 \times 10^{-4}$ cm$^{-3}$ based on study by Jenkins[214]). [d]"Low-metal" value (Lee et al.[215]). [e]Solar value divided by ~4 (based on study by Snow et al.[216]).

The so-called "low-metal" elemental abundances have been adopted (see Table 3). The abundances of C, N, and O are the values derived in the diffuse cloud ζ Oph by Jenkins,[214] which result in a C/O abundance ratio of 0.55. In the case of the second-row elements S, Si, and P, the metals Na, Mg, and Fe, and the halogen Cl, we have adopted the abundances from Lee et al.,[215] which are based on the original low-metal values proposed by Graedel et al.[133] These latter authors reduced the abundances of heavy elements other than C, N, and O by a factor of 100 with respect to the values observed in diffuse clouds in an attempt to take into account the fact that refractory elements are likely to be more depleted in dense clouds than in the diffuse medium. Insights into the depletion behavior of the different elements can be obtained from the comprehensive analysis carried out by Jenkins.[214] In this study elemental abundances were analyzed for more than 200 lines of sight probing diffuse media with average densities of H atoms in the range of 0.01–10 cm$^{-3}$, and it was found that as the density of the medium increases indeed elements such as S, Si, P, Cl, Mg, and Fe are severely depleted, while O is less severely depleted, and C and N are almost not depleted. Interestingly, if the results of this study are extrapolated to a density of hydrogen nuclei of $2 \times 10^4$ cm$^{-3}$, typical of dense clouds, the elemental abundances derived are not far from the low-metal values originally proposed by Graedel et al.[133] In the case of fluorine, which is also included in our model, we have assumed a depletion factor of ~4 with respect to the solar value, at the high edge of the range of values found by Snow et al.[216] for diffuse clouds. Some studies have proposed that the gas-phase C/O abundance ratio could be higher than unity in dark clouds with a rich content of carbon chain molecules.[99,137,169] To explore this possibility, we have also considered a carbon-rich case (C/O = 1.4) by reducing the elemental abundance of O relative to H to a value of $1.3 \times 10^{-4}$. For a broader discussion on the depletion of elements in cold dense clouds, we refer to the study by Hincelin et al.[99] Once the elemental abundances are chosen, it is customary to assume that initially all hydrogen is in the form of H$_2$ and the rest of the elements are in atomic form, either neutral or ionized depending on whether their ionization potential is above or below that of H. Here we have adopted this choice of initial abundances (see Table 3), which is however a matter of much debate.[128,217–219] Indeed, chemical abundances at any time before the steady state is reached depend to a large extent on the initial composition (also at the steady state if bistability occurs). The above choice is supposed to be representative of the chemical state of diffuse clouds, the likely precursor to the dense cloud stage. However, the evolution in the physical conditions from the diffuse to the dense state is not taken into account in pseudo-time-dependent chemical models. Moreover, diffuse clouds may have a wide variety of compositions, at one end containing mostly atomic rather than molecular hydrogen and at the other containing a variety of simple molecules having non-negligible abundances.[220–222] In any case, whatever the choice of initial abundances, it may be argued that it is somewhat arbitrary unless the dynamic evolution of the cloud is taken into account.

Our intention is to evaluate the ability of state of the art chemical networks to account for the molecular abundances observed in chemically rich dark clouds such as TMC-1 and L134N. We have therefore used the latest chemical network release of the databases KIDA (kida.uva.2011)[70,93] and UMIST (RATE12).[94,95] The two adopted networks are restricted to gas-phase processes, because here we aim at concentrating on the successes and failures of gas-phase chemistry. The only process related to dust grains included is the formation of $H_2$ on grain surfaces, for which we have adopted the canonical value $R_f = 3 \times 10^{-17}$ cm$^3$ s$^{-1}$, derived by Jura[223] for the diffuse interstellar medium.[224] The two chemical networks have similar numbers, the KIDA network containing 471 gas species linked by 6072 reactions and the UMIST network containing 6173 reactions involving 468 gas species. For a detailed description of the species and chemical processes included in the reaction networks, we refer to the original papers that describe the databases.[70,94] In the next sections, we analyze in detail the results of the chemical model and compare them with the abundances observed in the chemically rich dark clouds TMC-1 and L134N (see Table 4).

3.2.1. Oxygen Chemistry.

According to the behavior of the calculated abundance as a function of time, it is useful to divide species into two general categories. On one side, we have those species whose abundance peaks well before the steady state is reached, typically at $10^5$–$10^6$ yr, which we call "early- type" species. On the other, there are those species whose abundance increases with time, reaching its maximum value at the steady state, which are referred to as "late-type" species. By looking to the general behavior of oxygen-containing molecules (see Figure 2), we can clearly appreciate that molecules such as CO, HCO$^+$, and O$_2$ belong to this latter group while others such as H$_2$CO, CH$_2$CO, and CH$_3$CHO belong to the first category.

One of the major successes of pseudo-time-dependent chemical models of cold dense clouds is the prediction that CO is the most abundant molecule after H$_2$, with an abundance relative to that of H$_2$ of about $10^{-4}$, in good agreement with observed values. The high abundance of CO is also a strong indication of the extremely large departure from thermochem- ical equilibrium that exists in the interstellar medium. At temperatures as low as 10 K, chemical equilibrium predicts that most of the carbon will form CH$_4$ while oxygen will be present as H$_2$O, the abundance of CO being vanishingly small. In dark clouds, carbon monoxide is not formed by a single pathway but through many reactions, which have already been described in the literature.[131] We note that, although CO may be seen as a late-type species, it reaches its steady-state maximum abundance at a relatively early time of ~$10^5$ yr.

Table 4. Molecular Abundances (Relative to That of $H_2$) Observed in TMC-1 and L134N[a]

| | TMC-1 | L134N | | TMC-1 | L134N |
|---|---|---|---|---|---|
| OH | 3(-7)[221] | 7.5(-8)[221] | $NH_3$ | 2.45(-8)[190] | 5.9(-8)[191] |
| $H_2O$ | <7(-8)[222] | <3(-7)[222] | $N_2H^+$ | 2.8(-10)[190] | 6.1(-10)[191] |
| $O_2$ | <7.7(-8)[223]* | <1.7(-7)[223]* | CN | 7.4(-10)[190] | 4.8(-10)[191] |
| CO | 1.7(-4)[190] | 8.7(-5)[191] | HCN | 1.1(-8)[190] | 7.3(-9)[191] |
| $HCO^+$ | 9.3(-9)[190] | 7.9(-9)[191] | HNC | 2.6(-8)[190] | 2.6(-8)[191] |
| $H_2CO$ | 5(-8)[224] | 2(-8)[221] | $H_2CN$ | 1.5(-11)[225] | – |
| $CH_3O$ | <1(-10)[226] | – | $HCNH^+$ | 1.9(-9)[227] | <3.1(-9)[221] |
| $CH_3OH$ | 3.2(-9)[190] | 5.1(-9)[191] | $CH_2CN$ | 4(-9)[224] | <1(-9)[221] |
| HCOOH | – | 3(-10)[221] | $CH_3CN$ | 6(-10)[224] | <1(-9)[221] |
| $C_2O$ | 6(-11)[224] | – | $CH_2CHCN$ | 1(-9)[224] | <1(-10)[221] |
| $CH_2CO$ | 6(-10)[224] | <7(-10)[221] | $C_3N$ | 6(-10)[224] | <2(-10)[221] |
| $CH_3CHO$ | 6(-10)[228] | 6(-10)[221] | $HC_3N$ | 1.6(-8)[229] | 4.3(-10)[191] |
| $C_3O$ | 1(-10)[224] | <5(-11)[221] | $HC_2NC$ | 2.9(-10)[230] | – |
| HCCCHO | 8(-11)[224] | – | $HNC_3$ | 3.8(-11)[231] | – |
| CH | 2.0(-8)[221] | 1.0(-8)[221] | $HC_3NH^+$ | 1(-10)[224] | – |
| $C_2H$ | 7.2(-9)[190] | 2.3(-9)[191] | $CH_3C_3N$ | 8(-11)[224] | – |
| $c$-$C_3H$ | 1.03(-9)[232] | 4.3(-10)[233] | $CH_2CCHCN$ | 2.0(-10)[234] | – |
| $l$-$C_3H$ | 8.4(-11)[232] | 1.25(-10)[233] | $C_5N$ | 3.1(-11)[235] | – |
| $c$-$C_3H_2$ | 5.8(-9)[232] | 2.08(-9)[233] | $HC_5N$ | 4(-9)[224] | 1(-10)[221] |
| $l$-$C_3H_2$ | 2.1(-10)[232] | 4.2(-11)[233] | $CH_3C_5N$ | 7.4(-11)[236] | – |
| $CH_3C_2H$ | 6(-9)[221] | 5.9(-9)[233] | $HC_7N$ | 1(-9)[224] | <2(-11)[221] |
| $CH_2CHCH_3$ | 4(-9)[237] | – | $HC_9N$ | 5(-10)[224] | – |
| $C_4H$ | 7.1(-8)[238] | 1.77(-9)[233] | $HC_{11}N$ | 2.8(-11)[61] | – |
| $C_4H^-$ | <3.7(-12)[238] | <3.8(-12)[238] | NO | 2.7(-8)[239] | 2.0(-8)[240] |
| $H_2C_4$ | 7(-10)[224] | 3.1(-11)[233] | HNCO | 4(-10)[241] | 2.5(-10)[241] |
| $C_5H$ | 8(-10)[224] | <4.96(-11)[233] | HCNO | <1.4(-12)[241] | 5(-12)[241] |
| $CH_3C_4H$ | 1(-9)[224] | <6.7(-10)[233] | HOCN | <1.9(-12)[242] | 4(-12)[242] |
| $C_6H$ | 4.1(-10)[243] | <4.3(-11)[244] | $H_2S$ | <5(-10)[245] | 8(-10)[221] |
| $C_6H^-$ | 1.0(-11)[55] | <1.2(-11)[244] | CS | 2.9(-9)[190] | 9.9(-10)[191] |
| $H_2C_6$ | 4.7(-11)[246] | – | $HCS^+$ | 3(-10)[224] | 6(-11)[221] |
| $C_8H$ | 4.6(-11)[57] | – | $H_2CS$ | 7(-10)[224] | 6(-10)[221] |
| $C_8H^-$ | 2.1(-12)[57] | – | $C_2S$ | 7(-9)[224] | 6(-10)[221] |
| $CH_3C_6H$ | 3.1(-10)[247] | – | $C_3S$ | 1(-9)[224] | <2(-10)[221] |
| OCS | 2.2(-9)[248] | 3.5(-9)[248] | SO | 1.5(-9)[190] | 5.7(-9)[191] |
| NS | 8.0(-10)[249]* | 4.6(-10)[249]* | $SO_2$ | 3(-10)[250] | 2.6(-9)[191] |

* Abundance at position of $NH_3$ emission peak.
Notes: a(b) refers to a $\times 10^b$. Unless otherwise indicated abundances correspond to the positions TMC-1 $\alpha_{J2000}$ = $04^h 41^m 41^s.88$, $\delta_{J2000}$ = $+25°\ 41^m\ 27^s$ (cyanopolyyne peak) and L134N $\alpha_{J2000}$ = $15^h\ 54^m\ 06^s.55$, $\delta_{J2000}$ = $-2°\ 52^m\ 19^s$. Most abundances were derived from observed column densities adopting $N(H_2) = 10^{22}$ cm$^{-2}$ for both TMC-1[207] and L134N.[191,208]

References: [190] Pratap et al. 1997; [191] Dickens et al. 2000 (abundance at position C); [221] Ohishi et al. 1992; [222] Snell et al. 2000; [223] Pagani et al. 2003; [224] Ohishi and Kaifu 1998; [225] Ohishi et al. 1994; [226] Cernicharo et al. 2012; [227] Schilke et al. 1991; [228] Matthews et al. 1985; [229] Takano et al. 1998; [230] Kawaguchi et al. 1992; [231] Kawaguchi et al. 1992; [232] Fossé et al. 2001; [233] Turner et al. 2000; [234] Lovas et al. 2006; [235] Guélin et al. 1998; [236] Snyder et al. 2006; [237] Marcelino et al. 2007; [238] Agúndez et al. 2008; [239] Gerin et al. 1993; [240] Akyilmaz et al. 2007; [241] Marcelino et al. 2009; [242] Marcelino et al. 2010; [243] Bell et al. 1999; [244] Gupta et al. 2009; [245] Minh et al. 1989; [246] Langer et al. 1997; [247] Remijan et al. 2006; [248] Matthews et al. 1987; [249] McGonagle et al. 1994; [250] Cernicharo et al. 2011.

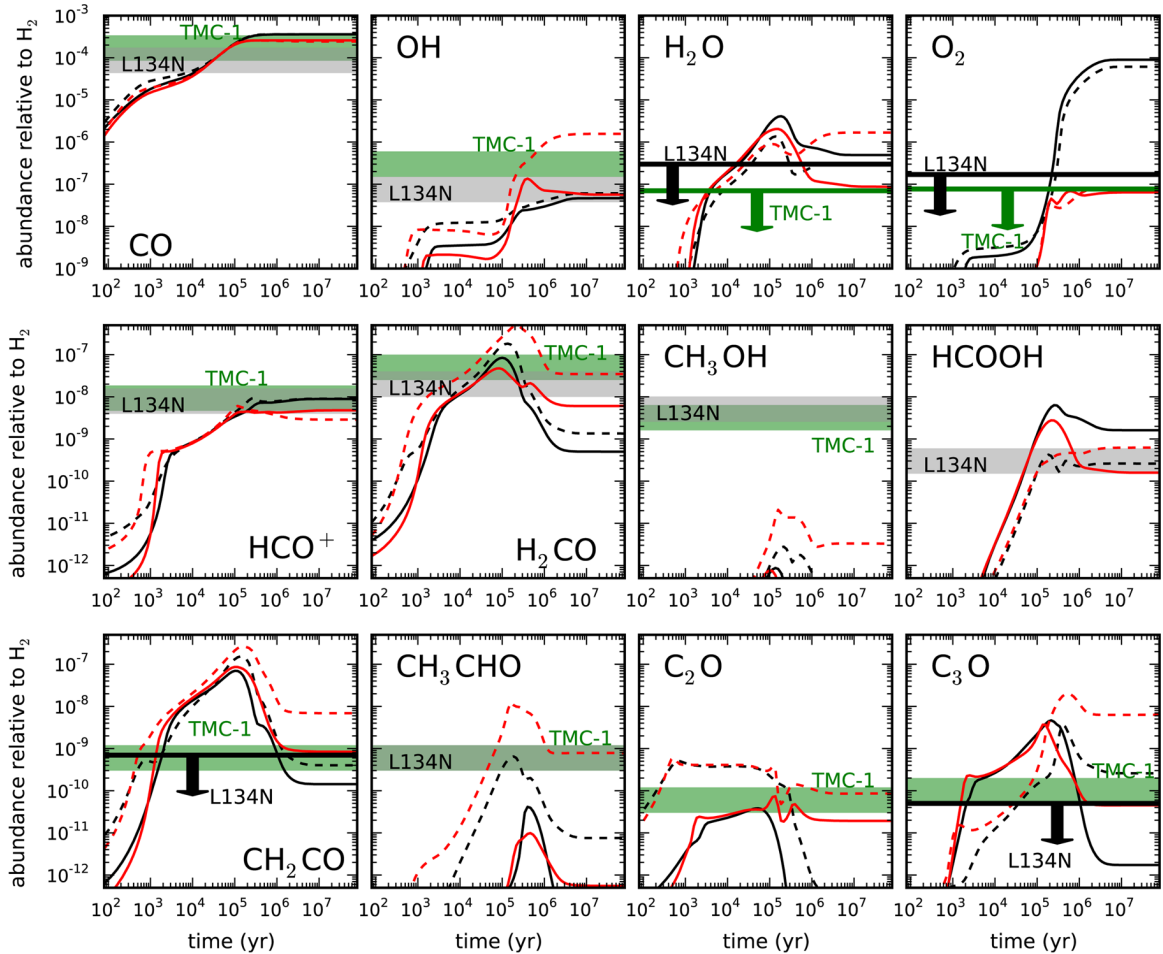

Figure 2. Abundances of oxygen-bearing molecules as a function of time, as calculated with the pseudo-time-dependent chemical model. Solid lines correspond to the results of the KIDA network and dashed lines to the model which uses the UMIST network. Black lines correspond to the "standard" oxygen-rich case and red lines to the carbon-rich case. Observed abundances (see Table 4) in TMC-1 (green horizontal bands) and L134N (gray horizontal bands) are plotted within a range of ±0.3 dex (in astronomy the term "dex" refers to the decimal logarithm), which corresponds to an optimistic uncertainty in the observed abundances. Observed upper limits are indicated by a downward arrow.

Most molecules observed in dark clouds are neutral, while just a few are ions, which is somewhat paradoxical taking into account that the chemistry is to a large extent dominated by ions. On one side, the fractional ionization in dark clouds is moderately low, in the range of $10^{-6}$–$10^{-9}$,[120,255] which results in low abundances for ions. On the other, the charge is not widespread among many ions but remains concentrated on just a few species. The positive charge is mostly carried by molecular ions, among them $HCO^+$ being the most abundant. Atomic ions such as $S^+$ and metals may also carry an important fraction of the positive charge depending on their degree of depletion. The formation of $HCO^+$ occurs at late times because it is mainly formed from the late-type species CO through the reaction

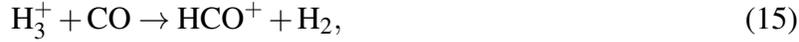

$$H_3^+ + CO \rightarrow HCO^+ + H_2, \qquad (15)$$

which is rapid at low temperatures, according to both experimental and theoretical investigations.[256] The main destruction reaction of HCO$^+$ is the dissociative recombination with electrons. HCO$^+$ being the most abundant molecular ion, it is easily observed in many dark clouds where it can be used to probe the degree of ionization.[255]

The case of O$_2$ is an interesting one. Molecular oxygen has long been predicted one of the most abundant molecules in cold dense clouds,[31,128,131,170] a fact that has motivated numerous sensitive searches for O$_2$ in the interstellar medium,[227,257] which have proven unsuccessful. Only recently O$_2$ has been finally detected in warm (~100 K) regions of the molecular cloud ϱ Oph[258,259] with an abundance relative to that of H$_2$ of about $5 \times 10^{-8}$. At this same level of abundance, upper limits have been derived in various cold dense clouds where O$_2$ has not been detected[227] (see Table 4). According to the pseudo-time-dependent gas-phase chemical model, O$_2$ is a late- type species which, under oxygen-rich conditions, reaches an abundance relative to that of H$_2$ above $10^{-5}$ after some $10^5$ yr (see Figure 2). It is mainly formed by the reaction

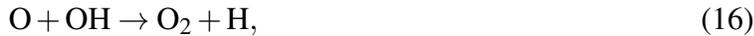

$$O + OH \rightarrow O_2 + H, \qquad (16)$$

for which the measured rate constant is about $3.5 \times 10^{-11}$ cm$^3$s$^{-1}$ in the temperature range of 39–142 K,[260] although some discussion exists on the possibility that the reaction may become significantly faster or slower around 10 K.[99,261,261] Clearly, the abundance of O$_2$ is severely overestimated by the chemical model unless very young cloud ages ($<10^5$ yr) or carbon-rich conditions are invoked.[182] Although any of these somewhat unusual requirements may pertain to some specific clouds, it seems unlikely that they can be generally extended to all dark clouds where molecular oxygen has been unsuccessfully searched for. Viti et al.[262] have proposed that the low observed abundance of molecular oxygen could be explained by chemical models presenting bistable solutions, in particular the HIP solution first introduced by Le Bourlot et al.[187] This solution, however, is also characterized by low molecular abundances for species such as CO.[171] Unless current gas-phase chemical networks miss some important data relevant for O$_2$, it seems that dust grain processes may hold the key to the low abundance of gas O$_2$. Indeed accretion onto dust grains at late times may considerably deplete O$_2$ from the gas phase. The molecules of O$_2$ adsorbed onto grains may, on one hand, be desorbed back to the gas phase and, on the other, be processed by grain surface reactions. The competition between these two processes would become crucial for the resulting abundance of gas O$_2$. Such processes are however not fully understood, and a variety of results concerning the gas-phase abundance of molecular oxygen are found by different models.[99,263,264] The processing of

O$_2$ on grain surfaces through successive hydrogenation reactions results in the formation of HO$_2$ and H$_2$O$_2$, whose further reaction with H atoms yields H$_2$O.[265,266] Interestingly, HO$_2$ and H$_2$O$_2$ have been recently observed in the gas phase of the warm molecular cloud ρ Oph,[267,268] where gas-phase O$_2$ was previously detected, which strengthens the grain surface O$_2$ chemical scheme.

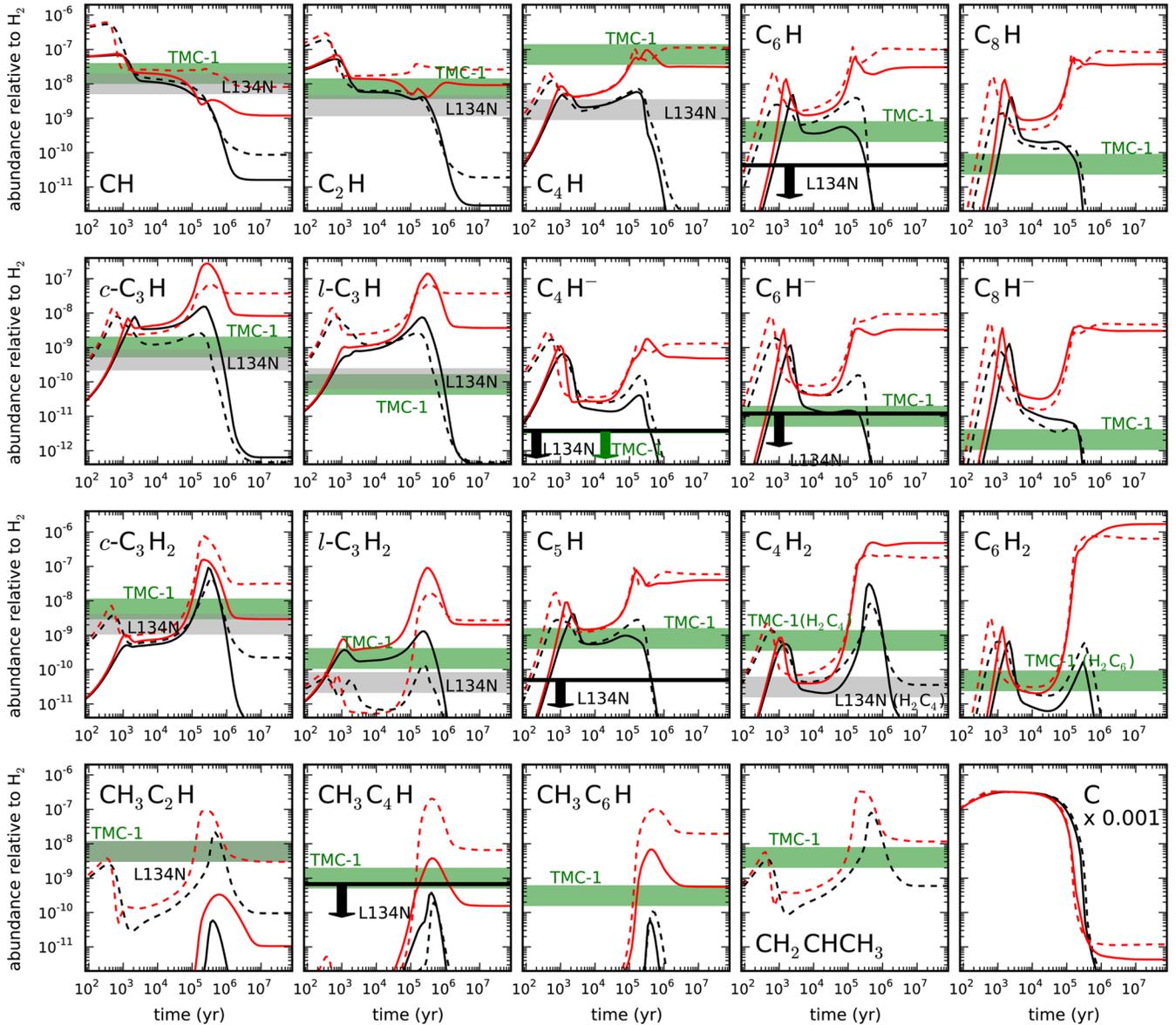

Figure 3. Same as Figure 2 but for hydrocarbons. For c-C$_3$H and l-C$_3$H the calculated abundances using the UMIST network are the same since this network considers just one isomer for this species. In the cases of C$_4$H$_2$ and C$_6$H$_2$ both the KIDA and UMIST networks consider just one isomeric form, while the observed abundances correspond to the carbenes H$_2$C$_4$ and H$_2$C$_6$ (the more stable polyyne isomers have no dipole moment and therefore cannot be detected through their rotational spectrum). Note that, in the bottom right panel, the abundance of C has been scaled down by a factor of 1000 for visualization purposes.

Water vapor is also predicted as a moderately abundant molecule, with an abundance relative to that of $H_2$ in the range of $10^{-6}$–$10^{-7}$ after some $10^4$ yr. The main formation route through neutral–neutral reactions being closed at 10 K (the successive reactions of O with $H_2$ have significant activation barriers), $H_2O$ is mostly formed through the following sequence of reactions involving ions:

$$H_3^+ + O \rightarrow OH^+ + H_2, \qquad (17)$$

$$OH^+ + H_2 \rightarrow H_2O^+ + H_2, \qquad (18)$$

$$H_2O^+ + H_2 \rightarrow H_3O^+ + H_2, \qquad (19)$$

$$H_3O^+ + e^- \rightarrow H_2O + H, \qquad (20)$$

The sequence begins with the protonation of atomic oxygen by $H_3^+$ and ends with the dissociative recombination of the molecular ion $H_3O^+$. Experiments have shown that this latter reaction yields water in 25% of the reactive collisions between $H_3O^+$ and an electron.[269] According to observations, however, the abundance of gas $H_2O$ is less than about $10^{-7}$ relative to that of $H_2$ in dark clouds such as TMC-1 and L134N (see Table 4). At this same level of abundance water vapor has been recently firmly detected for the first time in a starless core, in L1544 with an abundance of about $10^{-7}$ relative to that of $H_2$.[270] Water is expected to condense onto grains at late times and to be strongly bound to grain surfaces. Both gas–grain chemical models[36,173,263,264] and observations[271–273] indicate that water ice is the major constituent of grain mantles in dark clouds, the amount of water in the form of ice being much higher than that in the gas phase. Moreover, the moderately high abundance of water vapor derived in the starless core L1544 has been interpreted by Caselli et al.[270] as due to photodesorption of water ice molecules by UV photons induced by the impact of cosmic rays with $H_2$, the so-called Prasad–Tarafdar mechanism.

The case of $CH_3OH$ is clearly one of the major failures of current pseudo-time-dependent gas-phase chemical models of cold dense clouds. Methanol is commonly observed in many dark clouds with abundances relative to that of $H_2$ in the range of $10^{-8}$–$10^{-9}$. It has been postulated that it could be efficiently formed in the gas phase through the two-step process

$$CH_3^+ + H_2O \rightarrow CH_3OH_2^+ + h\nu, \qquad (21)$$

$$CH_3OH_2^+ + e^- \rightarrow CH_3OH + H, \qquad (22)$$

Recent measurements indicate that the rate constant of the radiative association is lower than $2 \times 10^{-12}$ cm$^3$ s$^{-1}$ at 50 K,[107] about 100–1000 times less than previous estimates,[274,275] while the dissociative recombination of protonated methanol leads to CH$_3$OH in just a small fraction of cases (3 ± 2%).[108] By using these updates, which are included in both KIDA and UMIST chemical networks, the calculated abundance of methanol remains various orders of magnitude below the observed value at any time (see Figure 2). Gas-phase chemistry not being able to account for the observed methanol in dark clouds, it seems that chemical reactions on grain surfaces together with some effective desorption process become the most likely formation mechanism. In the same vein, the recent detection of complex organic molecules such as dimethyl ether (CH$_3$OCH$_3$), methyl formate (HCOOCH$_3$), and the methoxy radical (CH$_3$O) in cold dense clouds[110,111,230] points toward grain surface chemistry as a much more important actor than previously believed.

3.2.2. Carbon Chemistry.

The chemistry of dark clouds, as in most interstellar regions, is a carbon-based chemistry, as indicated by the fact that most observed molecules are organic; i.e., they contain at least one carbon atom (see, e.g., Table 4). In pseudo-time-dependent models of cold dense clouds which use oxygen-rich conditions, the synthesis of hydrocarbons is usually a transitory phenomenon which takes place during the time at which neutral atomic carbon reaches its maximum abundance, which occurs earlier than $10^5$–$10^6$ yr for our chosen physical conditions (see Figure 3). At late times most of carbon goes into CO, and hydrocarbons show low steady-state abundances. One of the major differences found when carbon-rich conditions are adopted is that at late times the carbon excess not locked into CO can be used to form large hydrocarbons and carbon chain molecules, which maintain much higher abundances. The chemistry of hydrocarbons is dominated to a large extent by highly unsaturated (low H/C ratios) carbon chain molecules of the family of polyynes, such as the radicals C$_n$H. The butadiynyl radical (C$_4$H) stands out as one of the most abundant such molecules, especially in TMC-1. The formation of increasingly larger C$_n$H radicals in dark clouds occurs through ionic pathways of the types

$$C^+ + C_{n-1}H_2 \rightarrow C_nH^+ + H, \tag{23}$$

$$C + C_{n-1}H_2^+ \rightarrow C_nH^+ + H, \tag{24}$$

$$C_nH^+ + H_2 \rightarrow C_nH_2^+ + H, \tag{25}$$

$$C_nH_2^+ + e^- \rightarrow C_nH + H, \tag{26}$$

as well as by neutral–neutral reactions of the type

$$C + C_{n-1}H_2 \rightarrow C_nH + H, \tag{27}$$

In both cases the reaction of insertion of $C^+$ or $C$ becomes the crucial step to build an increasingly larger carbon skeleton. The rate constants of the reactions involved in the ionic pathway are relatively well constrained from extensive experiments.[91,276] Most of them have been carried out at room temperature and not for all relevant reactions, although ion–neutral reactions tend to be rather regular so that the extrapolation of rate constants to other analogous reactions and to lower temper- atures is usually not too risky. Most of the uncertainty related to ionic pathways concerns the lack of measurements of the product distribution in the dissociative recombination of hydrocarbon ions. Neutral–neutral reactions of the above type are less well constrained. The reaction of neutral carbon atoms with acetylene ($C_2H_2$) has been well studied[277–279] and found to be rapid at low temperatures, which suggests that analogous reactions are probably rapid as well.

Hydrocarbons observed in dark clouds are highly unsaturated and tend to be linear because of the alternating triple and single C–C bonds which characterize the backbone of polyynes. There are, however, some ring molecules, as in the case of $C_3H_2$ and $C_3H$, for which a cyclic isomer and a linear isomer are observed. Cyclopropenylidene (c-$C_3H_2$) is a stable three-membered ring which, in dark clouds such as TMC-1 and L134N, is observed to be 10–40 times more abundant than the linear carbene isomer propadienylidene (l-$C_3H_2$).[236,237] In the case of $C_3H$, observations of TMC-1 and L134N indicate that the cyclic form (c-$C_3H$) is also more abundant, by a factor of 4–12, than the linear isomer (l-C H).[236,237] The synthesis of these chain and ring hydrocarbons in cold dense clouds is supposed to occur from the precursor ion $C_3H_3^+$, for which a cyclic isomer (c-$C_3H_3^+$) and a linear isomer (l-$C_3H_3^+$) are possible. Both isomers are mainly formed via the radiative association of $C_3H^+$ and $H_2$:

$$C_3H^+ + H_2 \rightarrow C_3H_3^+ + h\nu, \tag{28}$$

In this reaction, equal amounts of c-$C_3H_3$ and l-$C_3H_3$ would be formed, according to theoretical calculations.[280] The cyclic species are then formed by the dissociative recombination of c-$C_3H_3^+$

$$\begin{aligned} c-C_3H_3^+ + e^- &\rightarrow c-C_3H_2 + H, \\ &\rightarrow c-C_3H + H_2, \end{aligned} \tag{29}$$

while l-$C_3H_2$ and l-$C_3H$ are formed by the dissociative recombination of the linear ion l-$C_3H_3^+$

$$\begin{aligned} l-C_3H_3^+ + e^- &\rightarrow l-C_3H_2 + H, \\ &\rightarrow l-C_3H + H_2, \end{aligned} \tag{30}$$

where it is supposed that the carbon skeletal structure is not altered during the dissociative recombination. Laboratory experiments have established that in the dissociative recombination of $C_3H_3^+$ species containing three carbon atoms accounts for 90.7% of the products, the remaining 9.3% corresponding to molecules with two carbon atoms[281] (channel not shown in reaction 30). In these experiments, however, it was not possible to determine either the amount of cyclic and linear $C_3H_3^+$ present in the reagent sample or the relative contributions of the product channels leading to $C_3H_2$ and $C_3H$. Laboratory measurements have also found that the dissociative recombination proceeds about 7 times faster for c-$C_3H_3^+$ than for l-$C_3H_3^+$.[282] The measurements were carried out in the 172–489 K temperature range, where little temperature dependence of the rate constants was observed, in disagree- ment with simple theory. This latter finding, if confirmed, together with the fact that c-$C_3H_3^+$ is less reactive with neutral species than l-$C_3H_3^+$,[91,276] may help to explain the cyclic-to-linear abundance ratios found in dark clouds. The situation, however, may be more complicated if additional reactions become important. For example, the reaction of $C_3H^+$ and $H_2$ may also yield, apart from the radiative association channels in reaction 28, c-$C_3H_2^+$ + H as products, although there are contradictory results[280,283] which do not allow whether the latter is an open channel to be concluded. Moreover, neutral– neutral reactions, such as C + $C_2H_2$,[277–279] and reactions involving negative ions, such as $C_3^-$ + H and $C_3H^-$ + H,[284] may also contribute to an important extent to the formation of the cyclic and linear isomers of $C_3H$ and $C_3H_2$.

The observation of negatively charged molecules has been one of the major recent surprises concerning the chemistry of cold dense clouds. Molecular anions are mainly formed by the radiative attachment of an electron to a neutral species, a process that may be schematically described as

$$A + e^- \rightarrow A^- + h\nu, \tag{31}$$

where A is any neutral species. This reaction is usually exothermic and may be very rapid if the neutral species has a high electron affinity and a moderately large size.[285] These requirements are easily fulfilled by large radicals, as is the case for the carbon chains $C_nH$ which lead to their anionic counterparts $C_nH^-$. On the other hand, negative ions are quite reactive species which can undergo reaction with atoms and neutralization with positive ions. In cold dense clouds the major destruction process of anions is probably the reaction with hydrogen atoms:

$$A^- + H \rightarrow AH + e^-, \tag{32}$$

This reaction is rapid according to laboratory measurements.[284,286] If the chemistry of molecular anions is restricted to the above two reactions, then it is possible to use the observed anion-to-neutral abundance ratio to derive the fractional abundance of electrons in dark clouds, provided the fractional abundance of H atoms is known.[58] Of course, it is also necessary to know the rate constant of the reaction of the negative ion with H, usually well constrained by experiments,[284,286] and of the electron attachment reaction, which is difficult to measure in the laboratory and must be theoretically calculated.[287] Such calculations seem to be fine for the electron attachment to large species such as $C_6H$ and $C_8H$, but are highly uncertain for smaller species. For example, in the case of the reaction of electron attachment to $C_4H$, there seems to be a severe problem since the calculated rate constant[287] results in a $C_4H^-/C_4H$ abundance ratio which is about 2 orders of magnitude higher than the values inferred from observations.[60,242,288]

Although most hydrocarbons observed in dark clouds are highly unsaturated, some partially saturated molecules are also observed. Among them, there are various methylpolyynes $CH_3C_nH$ (n = 2, 4, 6). In gas-phase chemical models of cold dense clouds, methylpolyynes are synthesized from the dissociative recombination of their protonated counterpart, which in turn is formed by the ion–neutral reaction

$$CH_4 + C_nH_2^+ \rightarrow C_{n+1}H_5^+ + H. \tag{33}$$

This chemical scheme is based on experimental grounds for methylacetylene (n = 2) but not for larger methylpolyynes. The reaction between $CH_4$ and $C_2H_2^+$ has been studied in the laboratory and found to be rapid,[91,276] while the dissociative recombination of $C_3H_5^+$ is known to yield molecules with three carbon atoms in 86.7% of cases,[281] although neither the absolute rate constant nor the distribution of the different molecules with three carbon atoms is known, and some uncertainties remain regarding the various possible isomers. The synthetic routes of large molecules are in general much more uncertain than for smaller species because there are more reactions involved in their buildup, but also because multiple isomers are possible and detailed information on the different isomers is difficult to obtain and is therefore often lacking in current chemical networks.

Among the partially saturated hydrocarbons observed in dark clouds, propylene ($CH_2CHCH_3$) stands out as the one with the highest degree of saturation.[241] A gas-phase route to form propylene in cold interstellar clouds has been recently proposed,[289] which involves two consecutive reactions of radiative association between a hydrocarbon ion and $H_2$. Starting from the propargyl ion l-$C_3H_3^+$, this sequence would form the highly saturated ion $C_3H_7^+$, whose dissociative recombination could form propylene if the H elimination exit channel is a sizable one. This route is included in the UMIST chemical network, which predicts an abundance of propylene in good agreement with the observed value (see Figure 3). The proposed mechanism must however be taken with caution, given the high degree of saturation of $CH_2CHCH_3$ and previous failures of similar mechanisms proposed to explain the formation of highly hydrogenated molecules such as methanol (see the previous section). Moreover, a recent study based on a combination of experiments and accurate ab initio calculations indicates that the hypothetical radiative associations leading to $C_3H_7^+$ are not efficient at interstellar temperatures.[290] Grain surface reactions could thus be an alternative if nonthermal desorption processes are able to bring $CH_2CHCH_3$ to the gas phase.

3.2.3. Nitrogen Chemistry.

Apart from hydrogen and helium, nitrogen is probably the element that is least severely depleted from the gas phase onto dust grains in different types of interstellar regions, from the diffuse medium[214] to dense starless cores.[16,291–293] This fact, together with the relatively rich chemistry of this element, explains the large number of N- containing molecules observed in cold dense clouds (see Figure 4). According to our gas-phase pseudo-time-dependent chemical model, nitrogen is mostly in atomic form at times earlier than about $10^6$ yr (obviously, the choice of initial abundances is to a large extent responsible for this), while at later times $N_2$ becomes the most abundant N-containing species. Contrary to most molecular species in cold cores, the synthesis of $N_2$ out of atomic nitrogen involves only neutral– neutral reactions. Two different pathways have been identified:

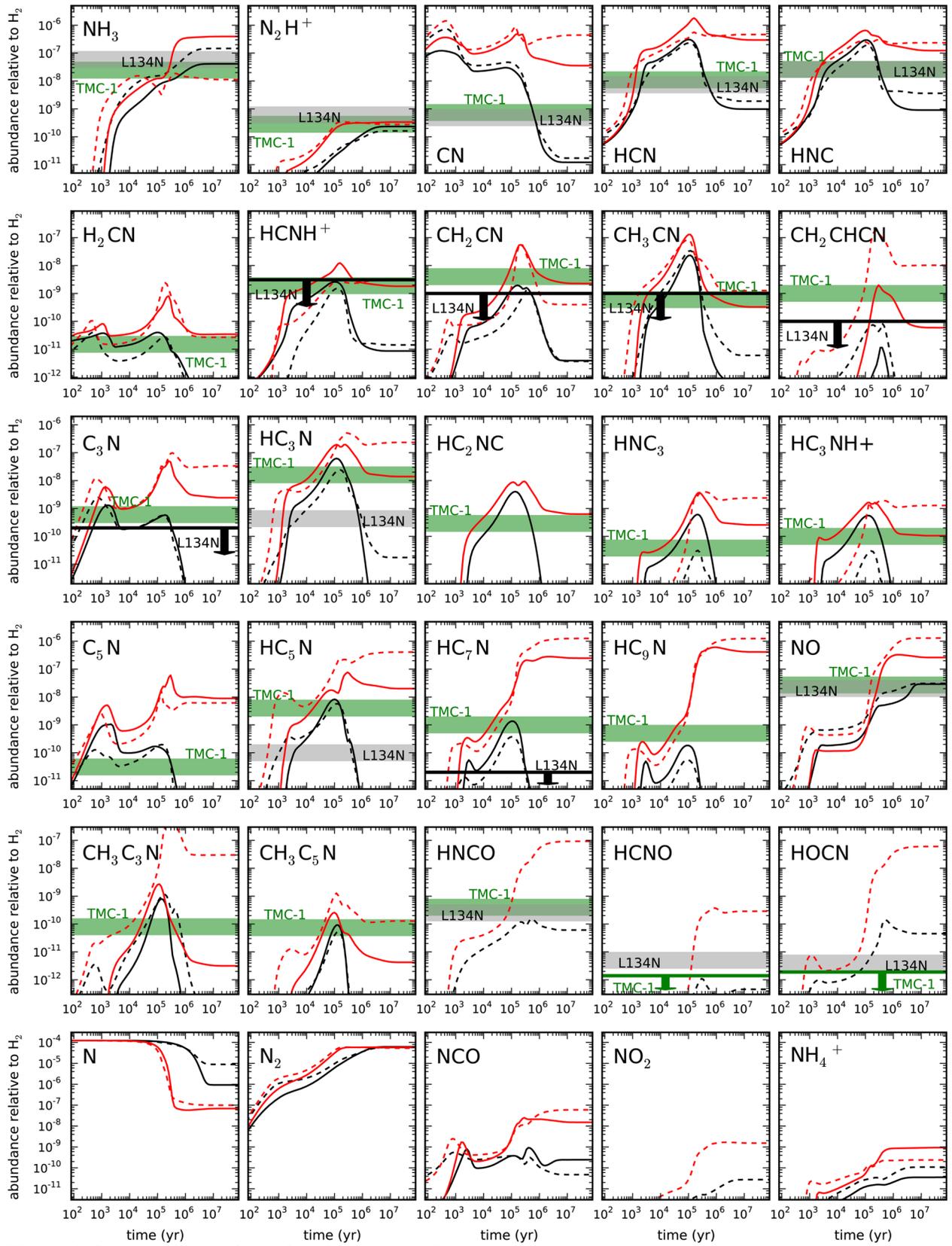

Figure 4. Same as Figure 2 but for nitrogen-bearing molecules.

$$N + OH \rightarrow NO + H, \tag{34}$$

$$N + NO \rightarrow N_2 + O, \tag{35}$$

and

$$N + CH \rightarrow CN + H, \tag{36}$$

$$N + CN \rightarrow N_2 + C. \tag{37}$$

Whether $N_2$ is the main reservoir of nitrogen in cold dense clouds is a matter of much controversy, and other forms such as atomic or $NH_3$ ice have been proposed as the major nitrogen species.[294,295] Unfortunately, the subject cannot be easily settled since molecular nitrogen has no dipole moment and cannot be observed through its rotational spectrum in dark clouds. In any case, $N_2$ is an important species in the gas-phase nitrogen network as the precursor of important molecules such as $NH_3$ and $N_2H^+$. The synthesis of ammonia in cold dense clouds occurs through a sequence of reactions similar to those which form water:

$$N_2 + He^+ \rightarrow N + N^+ + He, \tag{38}$$

$$N^+ + H_2 \rightarrow NH^+ + H, \tag{39}$$

$$NH^+ + H_2 \rightarrow NH_2^+ + H, \tag{40}$$

$$NH_2^+ + H_2 \rightarrow NH_3^+ + H, \tag{41}$$

$$NH_3^+ + H_2 \rightarrow NH_4^+ + H, \tag{42}$$

$$NH_4^+ + e^- \rightarrow NH_3 + H, \tag{43}$$

This sequence starts with the fragmentation of $N_2$ and formation of $N^+$ ions through the reaction with $He^+$, which in turn is formed by cosmic-ray ionization of He, followed by successive reactions with $H_2$ which end with the formation of the $NH_4^+$ ion, whose dissociative recombination with electrons finally yields $NH_3$ as the dominant product.[296,297] Reaction 39 has a small energy barrier of about 1.4 kJ mol$^{-1}$ (170 K if expressed as a temperature) when $H_2$ is in its ground para state[298] but becomes feasible at temperatures around 10 K if the ortho-to-para $H_2$ ratio exceeds a value of about $10^{-4}$, which is likely to be the case in cold dense clouds where values around $10^{-3}$ are inferred.[154,156,157,299] The synthesis of the $N_2H^+$, pretty similar to that of $HCO^+$, occurs by direct protonation of $N_2$ from $H_3^+$:

$$H_3^+ + N_2 \rightarrow N_2H^+ + H_2. \qquad (44)$$

Both ammonia and $N_2H^+$ are late-type species (see Figure 4) because they are formed from $N_2$, which is also produced at late times. Interestingly, $NH_3$ and $N_2H$ have been observed to survive in the gas phase in dense cores of starless clouds (with densities of particles between a few $10^4$ and a few $10^6\,cm^{-3}$), where other molecules such as CO are substantially depleted from the gas phase due to freeze out onto dust grains.[291,292] The reasons for the less severe depletion of nitrogen-bearing molecules onto dust grains, compared to other families of molecules, are not completely clear. As concerns $N_2$ and CO, it is reasonable to expect that the more volatile character of the former would facilitate its removal from grain surfaces by whatever desorption process is at work in dense cores. Ammonia and $N_2H^+$ can therefore be used to trace the physical conditions, and the ionization degree in the case of $N_2H^+$, of high-density cores of dark clouds.[16,300]

Most nitrogen-containing molecules observed in cold dense clouds are organic in nature; i.e., they contain at least one carbon atom. This is the case for hydrogen cyanide (HCN) and its less stable isomer hydrogen isocyanide (HNC), which are among the most abundant N-containing organic molecules in dark clouds. The observed HNC/HCN abundance ratio is in the range of 0.54–4.5, which indicates that HNC is somewhat more abundant than HCN.[301] Both molecules are mainly formed by the dissociative recombination of the precursor ion $HCNH^+$:

$$\begin{aligned}HCNH^+ + e^- &\rightarrow HCN + H, \\ &\rightarrow HNC + H,\end{aligned} \qquad (45)$$

For this reaction, the branching ratios for the formation of each isomer have not been firmly established by experiments, although various studies indicate that HCN and HNC should be produced in equal amounts.[302–305] Equal branching ratios are therefore adopted in current chemical networks, which predict an HNC/HCN abundance ratio close to unity in dark clouds (in the range of 0.6–2 at times later than $10^5$ yr in our pseudo-time-dependent model). One possible reason for the slightly overabundance observed in dark clouds for HNC with respect to HCN could be related to the existence of a less stable isomer of the precursor ion, $H_2NC^+$, whose dissociative recombination is likely to result in the isocyanide rather than the cyanide form, unless there is a strong rearrangement of bonds. The chemistry of formation of the isomer $H_2NC^+$ is, however, very badly constrained. It could also happen that there is a differential destruction rate for HCN and HNC, although it seems difficult to imagine the more stable isomer HCN being more reactive than HNC. It has also been pointed out recently that HNC/HCN abundance ratios derived from observations should be corrected downward, from >1 to ~1, due to differences in the cross sections of rotational excitation by inelastic collisions between HCN and HNC.[306] It is worth noting that recent observations of HCN and HNC in starless clouds have found HNC/HCN abundance ratios close to unity by using the new set of collision rate coefficients,[307] which points toward equal branching ratios for the production of HCN and HNC in reaction 45.

As occurs with hydrocarbons, nitrogen-containing molecules observed in dark clouds are dominated to a large extent by unsaturated species of the family of cyanopolyynes, such as $HC_nN$ and $C_nN$ with odd values of n. These species can be synthesized by reactions involving ions as well as by reactions between neutrals. In the case of cyanopolyynes, one of the major formation routes involves, similarly to the case of HCN, the dissociative recombination of the corresponding protonated nitrile ion:

$$HC_nNH^+ + e^- \rightarrow HC_nN + H, \quad (46)$$

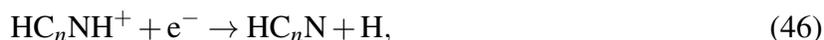

For this reaction, the rate constant and some information concerning the branching ratios have been obtained from experiments with the deuterated species $DC_3ND^+$.[308] Neutral–neutral reactions, mostly involving N atoms and hydrocarbon radicals, are also important in the synthesis of nitriles. For example, the reaction

$$N + C_nH \rightarrow C_nN + H, \quad (47)$$

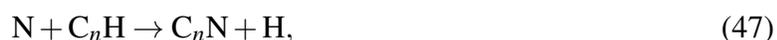

where n is odd and higher than 1, dominates the synthesis of the cyanopolyyne radicals $C_nN$. These types of reactions between an atom and a radical are however very difficult to study in the laboratory so that the rate constants used are mere guesses based on chemical intuition and related reactions.[140] Also, the recent discovery of molecular anions in the interstellar medium and the implementation of their chemistry in reaction networks has revealed that some reactions involving negative ions, such as

$$N + C_n^- \rightarrow C_nN + e^-, \quad (48)$$

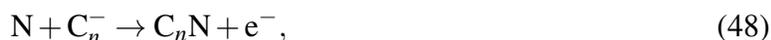

can be important in the synthesis of nitriles.[309] Reaction 48 has been studied experimentally for various values of n, and it has been found that it is rapid and that the associative electron detachment channel may indeed be a likely one.[284] The gas-phase pseudo-time-dependent chemical model predicts cyano- polyyne abundances at an early time of about $10^5$ yr which are in good agreement with the observed values in TMC-1 (see Figure 4). The general trend of decreasing abundance with increasing size is also well accounted for by the model. At late times, however, the abundances of large cyanopolyynes are severely underestimated if oxygen-rich conditions are used and overestimated under carbon-rich conditions. It is also interesting to note that, similarly to the case of HCN and HNC, the metastable isomers of cyanoacetylene, $HC_2NC$ and $HNC_3$ (with energies above that of the most stable isomer $HC_3N$ of 1.1 and 2.3 eV, respectively),[310] have been observed in dark clouds with sizable abundances.[234,235] Unlike the case of HCN and HNC, which have rather similar abundances, $HC_2NC$ and $HNC_3$ are about 50 and 400 times less abundant, respectively, than cyanoacetylene (see Table 4). These metastable isomers are likely to be formed in the dissociative recombination of protonated cyanoacetylene, although it is not clear whether their formation occurs from the most stable isomer $HC_3NH^+$, which would imply a strong rearrangement of bonds during the dissociative recombination, or whether they are formed from different isomers of the precursor ion with different backbone structures.

There are various cases of metastable isomers observed in cold dense clouds which serve to illustrate the large departure from thermochemical equilibrium that characterizes interstellar chemistry. A further case concerns isocyanic acid (HNCO) and its metastable isomers HOCN, HCNO, and HONC, which lie at energies of 1.1, 3.1, and 3.6 eV, respectively, above the most stable form HNCO. The two metastable isomers of lower energy (HOCN and HCNO) have been recently identified in dark clouds with abundances between 10 and 100 times lower than that of HNCO.[245,246] The chemistry of this family of isomers has been recently revised in light of the above-mentioned observational findings[246,311] and is implemented in the UMIST chemical network we have adopted here. According to our gas-phase pseudo-time-dependent chemical model, the synthesis of HNCO and HOCN is related to the dissociative recombination of the corresponding protonated ion having an NCO backbone structure, which in turn has several possible isomers:

$$H_2NCO^+ + e^- \rightarrow HNCO + H, \qquad (49)$$

$$\begin{aligned} HNCOH^+ + e^- &\rightarrow HNCO + H, \\ &\rightarrow HOCN + H, \end{aligned} \qquad (50)$$

$$H_2OCN^+ + e^- \rightarrow HOCN + H. \qquad (51)$$

The abundances calculated after about $10^5$ yr for HNCO and HOCN are on the order of the observed values in L134N when adopting oxygen-rich conditions, although the predicted HOCN/HNCO abundance ratio is on the order of unity while observations indicate that HNCO is substantially more abundant than HOCN. The isomer HCNO can also be formed by dissociative recombination of the corresponding protonated ion with a CNO backbone structure, although the main formation route involves the reaction

$$CH_2 + NO \rightarrow HCNO + H, \qquad (52)$$

which has been extensively studied (ref 311 and references therein). However, there are some uncertainties regarding whether HCNO is the main product. The abundance of HCNO calculated using the UMIST network and oxygen-rich conditions remains about 1–2 orders of magnitude below the observed value in L134N, although this result may depend to a large extent on the details of the adopted chemical network, since the calculations carried out by Marcelino et al.[246] and Quan et al.,[311] which used the Ohio State gas–grain network, predict much higher HCNO abundances in better agreement with observations. The chemistry of HNCO and related isomers is, in any case, very uncertain since most of the reactions involved in the ionic synthetic pathways are poorly constrained by experiments.

3.2.4. Sulfur Chemistry.

The presence of a good number of S-bearing molecules is an indication that sulfur is not extremely depleted in dark clouds, as seems to be the case for its second-row neighbors silicon and phosphorus. The true degree of depletion of sulfur in cold dense clouds is, however, very uncertain, and this uncertainty translates to the calculated abundances of S-containing species, which overall scale with the elemental abundance of S. According to the gas-phase pseudo- time-dependent chemical model, most of the sulfur is atomic at times earlier than about $10^6$ yr while at later times SO and $SO_2$ become the major reservoir of this element. It is interesting to note that the "late-type" species SO and $SO_2$ are indeed the most abundant observed S-bearing molecules in L134N, while in TMC-1 it is found that carbon monosulfide and the carbon chain molecules $C_2S$ and $C_3S$, which behave as "early-type" species (see Figure 5), have abundances similar or above those of sulfur oxides. The synthesis of sulfur monoxide relies exclusively on the reaction

$$S + O_2 \rightarrow SO + O, \qquad (53)$$

which is moderately rapid around room temperature according to extensive measurements,[312] although it has not been studied down to low temperatures. Sulfur dioxide is in turn formed from SO through the radiative association with oxygen atoms and the reaction with hydroxyl radicals. The abundances of SO and $SO_2$ therefore depend to a large extent on the abundance of $O_2$, for which there is a severe disagreement between observed and calculated abundances (see section 3.2.1 on oxygen chemistry).

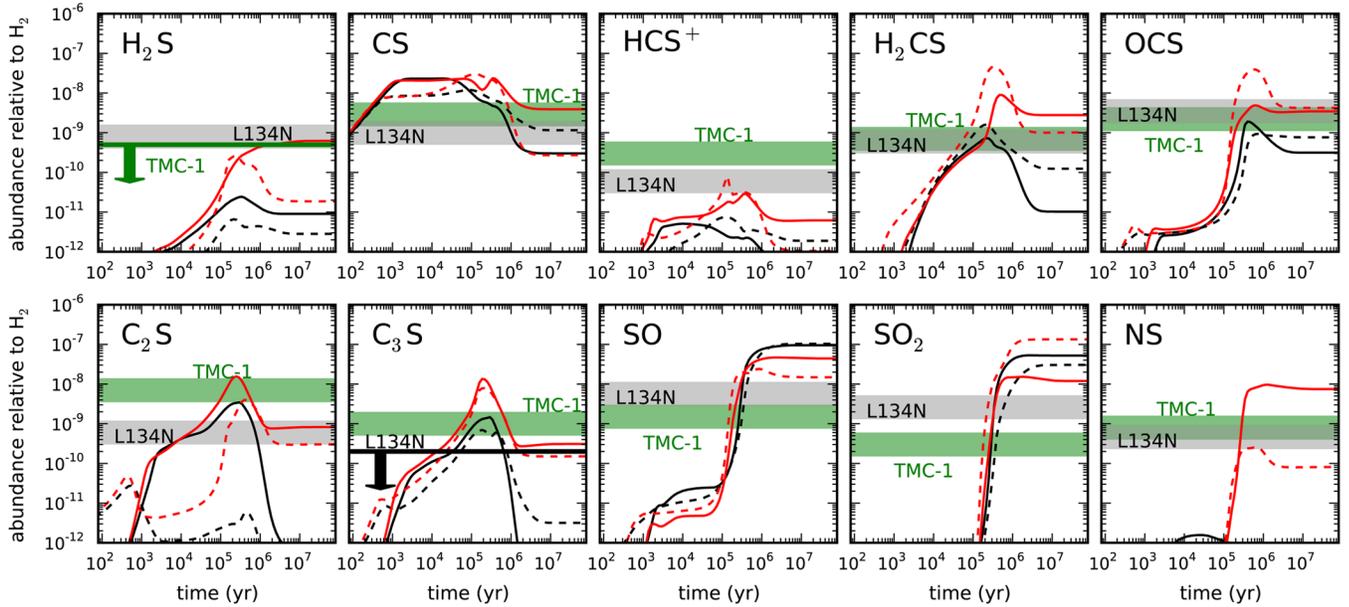

Figure 5. Same as Figure 2 but for sulfur-bearing molecules.

Various chemical routes have been proposed for the synthesis of S-containing carbon chain molecules in dark clouds.[313] In our gas-phase pseudo-time-dependent chemical model, the synthesis of $C_2S$ and $C_3S$ occurs primarily through reactions between $S^+$ and hydrocarbons, followed by the dissociative recombination of the organosulfur ion produced:

$$S^+ + C_nH_2 \rightarrow HC_nS^+ + H, \tag{54}$$

$$HC_nS^+ + e^- \rightarrow C_nS + H, \tag{55}$$

This pathway is supported by laboratory measurements of the reaction between $S^+$ and $C_2H_2$,[314] although data are lacking for dissociative recombinations of $HC_nS^+$ ions with n > 1. The abundances predicted for $C_2S$ and $C_3S$ at an early time between $10^5$ and $10^6$ yr are not far from the values observed in TMC-1 and L134N. However, the chemistry of these organosulfur molecules may be much more complex than the simple scheme given above. On one hand, other ion–neutral reactions such as

$$S^+ + C_nH \rightarrow C_nS^+ + H, \qquad (56)$$

$$C_nS^+ + H_2 \rightarrow HC_nS^+ + H, \qquad (57)$$

may also contribute to the formation of the precursor ion $HC_nS^+$, and on the other, neutral–neutral reactions could also play an important role, as indicated by various theoretical studies.[315,316] A further piece of this puzzle has been recently given by Sakai et al.,[317] who have found substantially different abundances for the isotopic species $^{13}CCS$ and $C^{13}CS$ in TMC- 1. This finding implies that the two carbon atoms of $C_2S$ are nonequivalent in its main formation reaction, which rules out reaction 54 with n = 2, yet it is the main route to the $HC_2S^+$ ion in our model. Clearly, further constraints from astronomical observations and chemical studies are needed to better understand the formation of these organosulfur molecules.

3.3. Comparison between the UMIST and KIDA Chemical Networks

For most molecules observed in TMC-1 and L134N, the calculated abundances using the UMIST and KIDA chemical networks are comparable, within 1 order of magnitude. There are, however, some molecules for which there are discrepancies larger than 1 order of magnitude in the calculated abundances. These occur mostly in the carbon-rich scenario for the molecules $CH_3CHO$ (see Figure 2), the methyl polyynes $CH_3C_2H$, $CH_3C_4H$, and $CH_3C_6H$ (see Figure 3), the N- bearing carbon chain molecules $CH_2CHCN$, $CH_3C_3N$, and $CH_3C_5N$ (see Figure 4), and the S-bearing molecules $H_2S$ and NS (see Figure 5). Overall, it is seen that carbon chain molecules usually reach lower abundances when using the KIDA network than when using the UMIST network. Although it is difficult to trace back the exact causes of the large differences found for some calculated abundances, a main source of these discrepancies is related to some neutral–neutral reactions which are included in the UMIST network and not in the KIDA chemical scheme. Mainly, these neutral–neutral reactions assist ion–neutral routes in the synthesis of carbon chain molecules. For example, the reaction between atomic oxygen and $C_2H_5$ is a main route to $CH_3CHO$ in the chemical scheme of UMIST, although it is not included in the KIDA network. Also, when using the UMIST network, there are various neutral–neutral reactions involving $C_2H$ and $C_4H$, which assist in the formation of $CH_3C_4H$ and $CH_3C_6H$, respectively, while others involving CN contribute to the formation of $CH_2CHCN$, $CH_3C_3N$, and $CH_3C_5N$, and these reactions are not considered in the KIDA network. Taking into account that the rate constants of many of these neutral–neutral reactions are mere estimations -- they have not been either measured in the laboratory or calculated by ab initio methods -- it is not clear which approach, either including or excluding them, is more correct. In any case the route toward a better convergence between different chemical networks involves necessarily the improvement of current chemical kinetics data.

## 3.4. Time Scales

Gas-phase pseudo-time-dependent chemical models of cold dense clouds can explain the observed abundances of a good number of molecules. Whether the capability of such a model to reproduce many of the observed abundances is somewhat fortuitous or a consequence of processes related to dust grains or cloud dynamics having little influence on the gas-phase chemical composition is a matter of much debate. Indeed, on one hand, a good agreement is only achieved at a given time, and it depends to a large extent on the elemental composition adopted, and on the other, we know that grain mantles are formed in these environments. Moreover, gas-phase pseudo- time-dependent chemical models also have some remarkable failures, as in the case of methanol. In spite of the above drawbacks, a useful exercise that permits an overall view of the goodness of this type of model is to evaluate the fraction of molecules which are correctly reproduced by the model at various times. Ideally, the comparison between observed and calculated abundances should take into account the un- certainties in both values. On one side, observed abundances are affected by telescope instrumental errors and by uncertainties arising from different assumptions during the process in which spectral line intensities are converted to fractional abundances. If these sources of uncertainty can be adequately constrained, then an error can be associated with an observed abundance, although the scatter of values found by different observational studies indicates that the uncertainties given are in many cases too optimistic. Typically, uncertainties in observed abundances may vary between a factor of 2 and 1 order of magnitude. On the other side, an error in the calculated abundances due to uncertainties in the reaction rate constants can also be estimated. Studies which have undertaken such a task[112,170] have found that the uncertainties in the calculated abundances increase with time (because of the accumulative effect of the propagation of uncertainties) and with the degree of complexity of the molecule (because more reactions are involved in the synthesis) and take values from somewhat less than a factor of 2, for simple molecules at early times, to slightly above 1 order of magnitude, for complex species at late times. For the sake of simplicity, here we merely consider that there is agreement between observed and calculated abundances if both values differ by less than 1 order of magnitude.

The fraction of molecules reproduced by the model using the above order of magnitude criterion is plotted in Figure 6 as a function of time. We have compared the abundances calculated using the KIDA and UMIST chemical networks and adopting oxygen-rich and carbon-rich conditions with the observed abundances in the chemically rich dark clouds TMC-1 and L134N (given in Table 4). In the case of TMC-1, under oxygen-rich conditions the best agreement between observed and calculated abundances (about 70% of reproduced molecules) occurs at an early time of $(2-3) \times 10^5$ yr. At late times (after $10^6$ yr) the agreement gets considerably worse mainly because the abundances calculated for complex species, such as large carbon chain molecules, fall well below the observed values. On the other side, the carbon-rich models show the opposite behavior, with a decline in the agreement at times between $10^5$ and $10^6$ yr (when large carbon chains are severely overpredicted by the model) and the best agreement reached outside this range of times (at $10^4$–$10^5$ yr and after $10^6$ yr). The better agreement found with the O-rich elemental abundances compared to C-rich conditions is in contradiction with previous chemical models.[140,170] The difference is related to the improvement of gas-phase rate coefficients included in the model, which demonstrates again the importance of such developments. As concerns L134N, the best agreement (above 70%) is reached under oxygen-rich conditions at a time of $(3–7) \times 10^5$ yr, somewhat later than in the case of TMC-1, while carbon-rich abundances result in a poorer agreement. We note that if large carbon chain molecules were considered in the analysis of L134N (abundance upper limits are however not available; see Table 4), the results would be qualitatively different at late times, with an improvement in the overall agreement under oxygen-rich conditions and a worsening when using C/O > 1.

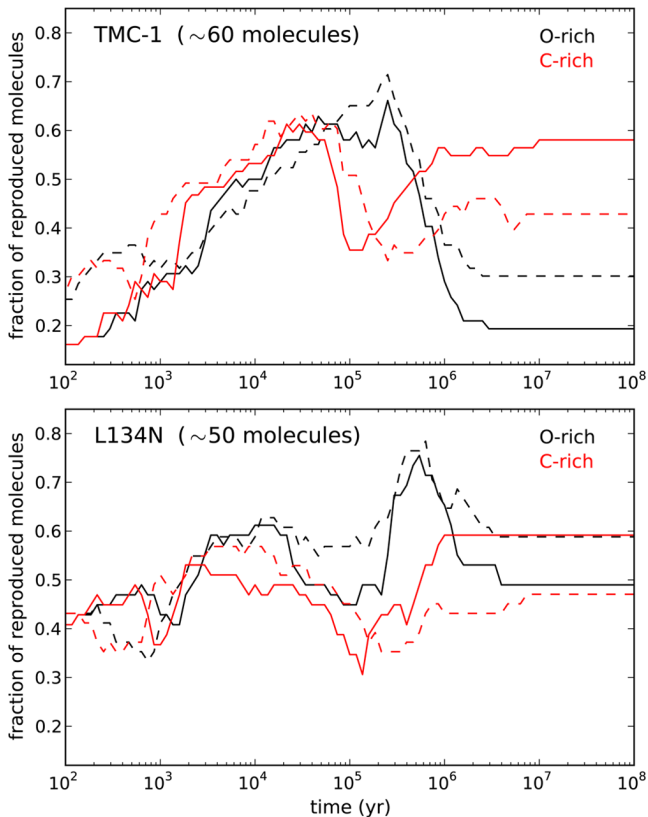

Figure 6. Fraction of reproduced molecules (for which the observed and calculated abundances differ by less than a factor of 10) as a function of time in TMC-1 (upper panel) and L134N (lower panel). Solid and dashed lines refer to the KIDA and UMIST chemical networks, respectively. Black lines correspond to the oxygen-rich case, while red lines correspond to the carbon-rich case.

From the behavior of the fraction of reproduced molecules as a function of time, we may naively infer that TMC-1 is either a carbon-rich source or a young cloud with an age of a few $10^5$ yr, while L134N is a somewhat older cloud with standard oxygen- rich conditions.

The possibility of extremely young ages ($<10^5$ yr) is probably too unrealistic taking into account that estimated lifetimes of cold dense clouds are in the range of $10^6$–$10^7$ yr.[105,106] Note that the presence of interstellar grains and gas–grain interactions changes the time of best agreement with observations.[83]

It is indeed tempting to use the time evolution of abundances calculated with a gas-phase pseudo-time-dependent model as a chemical clock to date cold dense clouds, although such a procedure may not be too meaningful for various reasons. The first is related to uncertainties in the chemical network itself, which is probably the major obstacle to make models more secure. Only a moderate fraction of the rate constants and product distributions of the many reactions included in current gas-phase chemical networks rely on experiments or theoretical calculations, most of them being mere guesses based on chemical intuition. Gas-phase chemical networks are also affected by a problem of lack of completeness, i.e., important reactions may be missing, and neglect processes related to dust grains, whose role in dark clouds may be more important than previously thought. Some of these problems may be alleviated by performing sensitivity analyses to quantify the uncertainties in the predicted abundances due to uncertainties in the reaction rate constants and by considering gas–grain exchange processes and chemical reactions of dust surfaces. A second reason is related to the pseudo-time-dependent approach itself, which neglects any physical evolution of the cloud and relies on a somewhat artificial choice of initial conditions, usually consisting of a gas of atoms (either neutral or ionized) and molecular hydrogen. Various studies have addressed this latter issue by considering the simultaneous evolution of the physical conditions and the chemical abundances, although there are important uncertainties in which is the most common scenario that leads to the formation of cold dense cores.[17,318,319]

In spite of the above shortcomings, gas-phase pseudo-time- dependent models of cold dense clouds can still provide some useful information regarding time scales. One of the most interesting outcomes of these models is that, unless carbon-rich conditions are assumed, the formation of complex molecules, especially large carbon chains, is predicted to be a relatively early and transient phenomenon that takes place at a so-called "early time" of a few $10^5$ yr. When comparing the model results with the abundances observed in chemically rich dark clouds such as TMC-1, the best agreement is obtained using these early-time abundances rather than steady-state values. An age of a few $10^5$ yr is probably too short with respect to the estimated lifetimes of cold dense clouds, in the range of $10^6$–$10^7$ yr,[105,106] and therefore, much debate exists on the meaning of such an early time. It is not clear whether a rich content of large carbon chain molecules in a dark cloud may be a sign of youth or evidence of carbon-rich gas or either of these possibilities. An important piece of the puzzle was given by Suzuki et al.,[320] who carried out astronomical observations of a sample of dark clouds and found that $C_2S$ shows a positive correlation with $HC_3N$ and $HC_5N$, while it shows no correlation with ammonia. These results were interpreted in terms of cloud evolution, and it was proposed that the $C_2S/NH_3$ abundance ratio could be a good indicator of the age of the cloud, $C_2S$ and $NH_3$ being abundant in the early and late stages of evolution, respectively. Marka et al.[321] carried out a comprehensive study of the $C_2S/NH_3$ abundance ratio in a selection of sources at different evolutionary stages and did not find any correlation, contrary to the predictions of chemical models.

The existence of a chemical differentiation among different dark cloud cores is well established from observations, and the idea that it could be due to different evolutionary stages has been developed in various studies. For example, it is known that the dark cloud TMC-1 has a fragmented structure with various dense cores, the most widely observed being the cyanopolyyne and ammonia peaks.[194,322] Hanawa et al.[323] have suggested that the cores are formed sequentially, an idea that has been explored by Howe et al.[217] in the framework of a core collapse model in which ages in the range of $(1-2) \times 10^5$ yr were deduced for each core (the cyanopolyyne peak being younger than the ammonia peak) by fitting the calculated column densities to the observed values. Pratap et al.[194] have suggested that the gradients in the molecular abundance observed along the TMC-1 ridge could be due to different evolutionary stages, but also to differences in the density or in the gas C/O abundance ratio. An alternative explanation proposed by Markwick et al.[183] links the chemical gradients observed in TMC-1 to the removal of icy grain mantles induced by the propagation of magnetohydrodynamic waves. From a theoretical point of view, the demonstration of a link between the chemical content and the evolutionary stage of a dark cloud is hampered by the lack of knowledge on how cold dense clouds form. The scenario could entail the gravitational collapse of a precursor diffuse cloud, although the role of processes such as shocks, turbulence, and magnetic fields remains poorly understood. Moreover, a scenario in which interstellar clouds would experience successive dynamical cycles between a diffuse and a dense phase has also been suggested.[142,206]

Taking into account that most chemical models of dark clouds base their reliability on a comparison with the abundances observed in very few sources, mostly in TMC-1, it would be desirable to extend the number of chemically rich dark clouds. Recent efforts to this end have been undertaken by Yamamoto and co-workers[324,325] in a search for carbon chain rich clouds which has yielded interesting candidates such as Lupus-1A, where the line intensities of long carbon chain molecules and their anionic counterparts are even brighter than those in TMC-1. It would be very interesting to characterize in detail the chemical composition of such objects to better understand the possible link between chemical composition and the evolutionary status of cold dense clouds.

3.5. Importance of Surface Chemistry

The simulations presented in section 3.2 and used for comparison with observations have been done ignoring the interactions with the surface of the grains. The freeze-out time scale of molecules at the typical densities of clouds is on the same order as the free-fall time of collapse of these clouds. Grain surface reactions and gas–grain interactions may then be necessary to consider to explain the observations in some of the sources. The main effect of these processes is the depletion of molecules from the gas phase. Using a pure gas-phase model, the major carbon-bearing species is CO, with a predicted abundance as high as the elemental abundance of carbon considered in the model (usually larger than $10^{-4}$ with respect to that of $H_2$). Including adsorption onto the grain surfaces diminishes the gas-phase abundance of CO after a few $10^5$ yr, closer to the observed abundance in L134N, for instance (see Table 3). Even though at low temperature the reactivity on the grain surfaces is low, some molecules can be formed and be reinjected into the gas phase by nondirect thermal processes, changing the gas-phase abundances as was proposed for methanol.[83] Considering both gas-phase and grain surface chemistries, Loison et al.[326] did the same comparison shown in section 3.2. The general conclusions on the age constraints are very similar, showing that, for the two considered clouds, grain surfaces do not influence much the gas-phase chemical composition.

# AUTHOR INFORMATION Corresponding Author

*E-mail: valentine.wakelam@obs.u-bordeaux1.fr. Notes

The authors declare no competing financial interest. Biographies

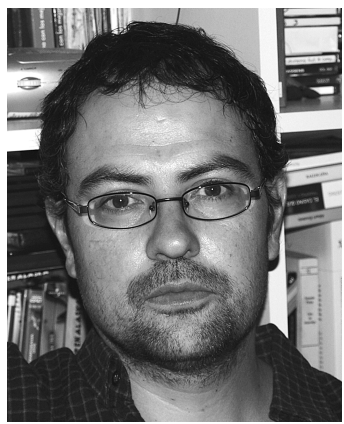

Marcelino Agúndez was born in Leon, Spain, in1978. He received his bachelor's degree in chemistry in 2003 from the Universidad de Valladolid (Spain) and his Ph.D. in 2009 from the Universidad Autońoma de Madrid (Spain) under the supervision of Prof. José Cernicharo. Since then, he has worked at the Observatoire de Paris-Meudon with Prof. Evelyne Roueff in the framework of a Marie Curie Intra-European Fellowship and at the Laboratoire d'Astrophysique de Bordeaux with Dr. Franck Selsis. His main research interests have focused on the interplay between chemistry and astrophysics covering various aspects such as radioastronomical observations, radiative transfer and chemical models of interstellar clouds, circumstellar sources, and atmospheres of exoplanets. He has actively contributed to the discovery of various molecules in space, among them molecular anions and phosphorus-bearing molecules, and has authored over 40 publications in the area of molecular astrophysics.

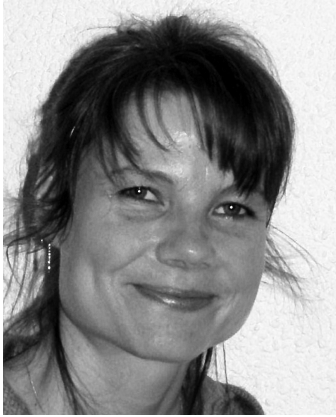

Valentine Wakelam was born in 1977 in Port of Spain, Trinidad and Tobago. After graduating from the French University of Guadeloupe (French West Indies), she obtained a Ph.D. degree in astrophysics at the University of Toulouse in France in 2004. Under the Ph.D. supervision of Cecilia Ceccarelli and Alain Castets, she studied sulfur chemistry in low mass star forming regions using both observational and theoretical approaches. As a postdoc, she worked for two years with Prof. Eric Herbst at The Ohio State University (United States) on the chemical modeling of dark clouds, in particular developing methods to study chemical model uncertainties. In 2006, she obtained a researcher position at the French CNRS institute and is now located at the Laboratoire d'Astrophysique de Bordeaux. Since then, she has been working on the chemical modeling of all the steps leading to the formation of planets starting from a diffuse medium.

## ACKNOWLEDGMENTS

We thank D. McElroy and T. J. Millar for kindly providing the latest chemical network release of the UMIST database prior to publication of the paper. M.A. acknowledges financial support from the European Research Council (ERC Grant 209622: E$_3$ARTHs). V.W.'s research is supported by the French Institut National des Sciences de l'Univers (INSU)/CNRS program PCMI (Physique et Chimie du Milieu Interstellaire) and the Observatoire Aquitain des Sciences de l'Univers.

centered at a wavelength of 550 nm, due to absorption and scattering by dust and gas. $A_V$ is measured in magnitudes, where 1 magnitude is a change in brightness of a factor of $100^{1/5} \approx 2.512$.

refers to a state in which the chemical composition corresponds to the minimum of the Gibbs energy of the system and must not be confused with the steady state obtained in the framework of chemical kinetics.